\documentclass[aps,showpacs,nofootinbib,amssymb,preprint]{revtex4-1}
\usepackage{graphicx,epsfig}
\usepackage{simplewick}
\usepackage{slashed}
\usepackage[retainorgcmds]{IEEEtrantools}
\usepackage{color}

\usepackage{dsfont} 
\usepackage{epstopdf}

\usepackage[pdftex]{hyperref}

\newcommand{\nn}{\nonumber}
\newcommand{\be}{\begin{equation}}
\newcommand{\ee}{\end{equation}}
\newcommand{\bea}{\begin{eqnarray}}
\newcommand{\eea}{\end{eqnarray}}

\def\siml{{\ \lower-1.2pt\vbox{\hbox{\rlap{$<$}\lower6pt\vbox{\hbox{$\sim$}}}}\ }}

\def\th{\theta}

\def\d{\delta}

\def\tr{\mathrm{Tr}}
\def\d{\mathrm{d}}
\def\p{\partial}
\def\e{\epsilon}
\def\de{\delta}

\def\sd{\slashed{\delta}}
\def\T{\mathcal{T}}
\def\H{\mathcal{H}}
\def\V{\mathcal{V}}

\def\half{{1\over2}}
\def\SE{Schr\"odinger equation}

\newcommand{\bE}{\begin{IEEEeqnarray}}
\newcommand{\eE}{\end{IEEEeqnarray}}

\def \beq  {\begin{equation}}
\def \eeq  {\end{equation}}
\def \beqar {\begin{eqnarray}}
\def \eeqar {\end{eqnarray}}
\begin{document}

\title{\boldmath 
The regularization and determination of the Yang-Mills vacuum wave functional in three dimensions at ${\cal O}(e^2)$
\unboldmath}
\author{Sebastian Krug and Antonio Pineda}
\affiliation{Grup de F\'\i sica Te\`orica, Universitat
Aut\`onoma de Barcelona, E-08193 Bellaterra, Barcelona, Spain}

\date{\today}

\begin{abstract}
\noindent
We complete the computation of the Yang-Mills vacuum wave functional in three dimensions at weak coupling with ${\cal O}(e^2)$ precision. We use two different methods to solve the functional Schr\"odinger equation. One of them generalizes to ${\cal O}(e^2)$ the method followed by 
Hatfield at ${\cal O}(e)$~\cite{Hatfield:1984dv}.  
The other uses the weak coupling version of the gauge invariant formulation of the Schr\"odinger equation and the ground state wave functional followed by Karabali, Nair, and Yelnikov \cite{Karabali:2009rg}. These methods need to be carefully regularized to yield correct results. This is done in this paper with full detail.
\end{abstract}
\pacs{11.10.Ef, 11.10.Kk, 12.38.-t} 
\maketitle

\section{Introduction}

In Ref.~\cite{Krug:2013yq} we computed the Yang-Mills vacuum wave functional in three dimensions at weak coupling to ${\cal O}(e^2)$. 
We used two different methods: (A) One extends to ${\cal O}(e^2)$ and to a general gauge group the computation performed in Ref.~\cite{Hatfield:1984dv} to ${\cal O}(e)$ for SU(2); (B) The other method is based on the weak coupling limit of the reformulation of the Schr\"odinger equation in terms of gauge invariant variables \cite{Karabali:1995ps,Karabali:1996je,Karabali:1996iu, Karabali:1997wk,Karabali:1998yq}, and on the approximated expression obtained in Ref. \cite{Karabali:2009rg} for the wave functional. 

In the comparison between both results we obtained almost complete agreement, except for one term. In Ref.~\cite{Krug:2013yq} 
we concluded that this discrepancy could be due to regularization issues, which had not been systematically addressed in that paper.
It is the purpose of this study to fill this gap and to provide with the complete expression of the Yang-Mills vacuum wave functional in three dimensions 
with ${\cal O}(e^2)$ precision for the first time.

The regularization of the Schr\"odinger equation and the vacuum wave functional in quantum field theories is a complicated subject. Whereas some formal aspects have been studied quite a while ago in Refs. \cite{Symanzik:1981wd,Luscher:1985iu}, there have not been many quantitative studies of the regularization of the 
Yang-Mills vacuum wave functional. In three dimensions, the most detailed analyses have been carried out using the method (B) (see, for instance, the discussion in Refs. \cite{Karabali:1997wk,Agarwal:2007ns}, in particular in the Appendix of the last reference). It is claimed in those references that the regularization has been completely taken into account. According to this, the result obtained in Ref.~\cite{Krug:2013yq} using the
method (B) (which corresponds to the weak coupling limit of the approximated expression obtained in Ref. \cite{Karabali:2009rg} for the wave functional) should be the correct one. We will actually see that this is not so and that the regularization procedure has to be 
modified to obtain the correct Yang-Mills vacuum wave functional in three dimensions at weak coupling. This produces a new contribution that 
has to be added to the result obtained in Ref.~\cite{Krug:2013yq}.

The result given in Ref.~\cite{Krug:2013yq} using method (A) was obtained without regularizing the functional Schr\"odinger equation. 
It directly works with the gauge variables ${\vec A}$, but it has the complication that the Gauss law constraint has to be implemented by hand. 
In the intermediate steps potentially divergent expressions were found, which, nevertheless could be handled formally (assuming that the symmetries of the classical theory survive) obtaining a finite result. In this paper we carefully regularize the computation using method (A). Out of this analysis a 
new contribution has to be added to the result obtained in Ref.~\cite{Krug:2013yq}.

The new results obtained for the Yang-Mills vacuum wave functional in three dimensions at weak coupling to ${\cal O}(e^2)$ with the methods (A) and (B) agree with each other. This is a strong check of our computations and of the regularization methods used in this paper. On the other 
hand our results imply that the weak coupling limit of the expression obtained in Ref. \cite{Karabali:2009rg} for the wave functional is not 
correct with ${\cal O}(e^2)$ precision (though it is at ${\cal O}(e)$).

This paper has a strong overlap with Ref.~\cite{Krug:2013yq}, from which we will borrow notation and several equations, and refer to it for more details (yet we will try to make this paper as self-contained as possible). Following that paper 
we will denote by $\Psi_{GL}[{\vec A}]$ the vacuum wave functional obtained using method (A) and $\Psi_{GI}[J]$ the one obtained using 
method (B).

The outline of the paper is the following: In Sec.~\ref{sec:Reg} we regularize the Schr\"odinger equation. In Sec.~\ref{sec:Hatfield} we compute the wave functional using the method (A) with ${\cal O}(e^2)$ precision. In Sec.~\ref{sec:KKN} we rewrite the regularized version of the Schr\"odinger equation obtained in Sec.~\ref{sec:Reg} in terms of the gauge invariant variables, and compute the wave functional using the method (B) with ${\cal O}(e^2)$ precision. We also discuss the reason why the Schr\"odinger equation used in Ref. \cite{Karabali:2009rg} is not sufficient to obtain the complete expression for the vacuum wave functional to ${\cal O}(e^2)$. Finally, a series of definitions and computations are relegated to the appendices.

\section{The regulated  \SE}
\label{sec:Reg} 
The Yang-Mills Lagrangian reads
\be
{\cal L}=-\frac{1}{4}G^{\mu\nu,a}G_{\mu\nu}^a
\,,
\ee
where 
\be
G_{\mu\nu}^a=\partial_{\mu}A_{\nu}^a-\partial_{\nu}A_{\mu}^a+ef^{abc}A^b_{\mu}A^c_{\nu}
\,,
\ee
$eG_{\mu\nu}=[D_{\mu},D_{\nu}]$, $D_{\mu}=\partial_{\mu} +eA_{\mu}$, $A_{\mu}=-iT^aA_{\mu}^a$, 
$G^{\mu\nu}=-iT^aG^{\mu\nu}_a$, $T^a$ are the $SU(N)$ generators (with $(T^a)_{bc}=-if^{abc}$ in the adjoint representation), and $[T^a,T^b]=if^{abc}T^c$. 

In the Schr\"odinger picture the ground state wave functional satisfies the time independent \SE:
\be
\H \Psi = E_0\Psi=0\,,
\ee
where in the last equality we have normalized the ground state energy to zero.

In the temporal gauge ($A_0=0$) we work with the spatial components only, ${\vec A}=(A_1, A_2)$, and we have the Hamiltonian\footnote{In the following we use the notation ($d=2$): $\int_x \equiv \int d^dx$, $\int_\slashed{k} \equiv \int \frac{d^2k}{(2\pi)^d}$, $\slashed{\delta}(\vec{k})\equiv (2\pi)^d\delta^{(d)}(\vec{k})$, and so on.}
\be
\H=\T+\V={1\over2}\int_x \left((\vec{E}^a(\vec x))^2+(B^a(\vec x))^2\right)
\ee
and the equal time commutators
\be
\left[E_i(\vec{x},t_0),A_j(\vec{y},t_0)\right] = i\de_{ij}\de^{(2)}(\vec{x}-\vec{y})\,,
\ee
where 
\be
B^a=\frac{1}{2}\epsilon_{jk}(\partial_jA_k-\partial_kA_j+e[A_j,A_k])^a
 = {\vec \nabla}\times\vec{A}^a +\frac{e}{2}f^{abc}\vec{A}^b\times\vec{A}^c
\,,
\ee
with ${\vec A}\times {\vec B} \equiv \epsilon_{ij}A_iB_j$, $\vec \nabla_i \equiv \partial_i=\partial/\partial x^i$ (for simplicity, we use the metric $\eta_{\mu\nu}=\mathrm{diag}(-1,+1,+1)$, so there is no sign difference between upper and lower spatial indices), and $B=-iT^aB^a$.\\
We realize the commutators by working in a representation where $A_i(\vec x)$ is diagonal and thus
\be
E_i(\vec x) =i\frac{\de}{\de A_i(\vec x)}\,,
\ee
and the \SE $ $ reads
\be
\frac{1}{2}\int_x\left(-\frac{\delta}{\delta \vec{A}^a(\vec{x})} \cdot \frac{\delta}{\delta \vec{A}^a(\vec{x})} + B^a(\vec{x}) B^a(\vec{x}) \right) 
\Psi = 0
\,. \label{GLH}
\ee

In order to regularize the kinetic operator we separate the points at which the differential operators act. As we want to preserve gauge invariance, we do this by introducing a Wilson line and a regulated delta function
\be
\delta_\mu(\vec x, \vec v)={\mu^2 \over \pi} e^{-(\vec{x}-\vec{v})^2\mu^2}\,,
\ee
such that after removing the regulator $\mu\to\infty$ one recovers the original expression:

\be
\mathcal{T}=-{1\over2}\int_x\frac{\delta}{\delta A_i^a(\vec x)}\frac{\delta}{\delta A_i^a(\vec x)} 
\longrightarrow 
\mathcal{T}_{reg}=-{1\over2}\int_{x,v} \delta_\mu(\vec x, \vec v)\frac{\delta}{\delta A_i^a(\vec x)} \Phi_{ab}(\vec x,\vec v) \frac{\delta}{\delta A_i^b(\vec v)}\,.
\label{Treg_middleString}
\ee
The first functional derivative also acts on the Wilson line, which ensures that the regulated kinetic operator is still hermitian.

The Wilson line is the path-ordered exponential of the gauge fields along a curve $\mathcal{C}$:
\be
\Phi(\mathcal{C};\vec x,\vec v)=\mathcal{P}e^{-e\int^{\vec x}_{\vec v}\d z^iA_i(\vec z)} = \mathcal{P}e^{-e\int_0^1\d s\, \dot{z}^i(s)A_i(\vec z(s))}\,,
\ee
where $\vec{z}(s)$ is the parametrization of $\mathcal{C}$.
The Wilson line transforms as
\be
\Phi(\mathcal{C};\vec x,\vec v) \to \left(g(\vec x)\Phi(\mathcal{C};\vec x,\vec v)g^\dagger(\vec v)\right)_{ab}
\ee
under gauge transformations
\be
A_i\to A_i^g=gA_ig^{-1}+\frac{1}{e}g\p_ig^{-1} \,.
\ee

\begin{figure}
\centerline{
\includegraphics[width=.49\textwidth,clip]{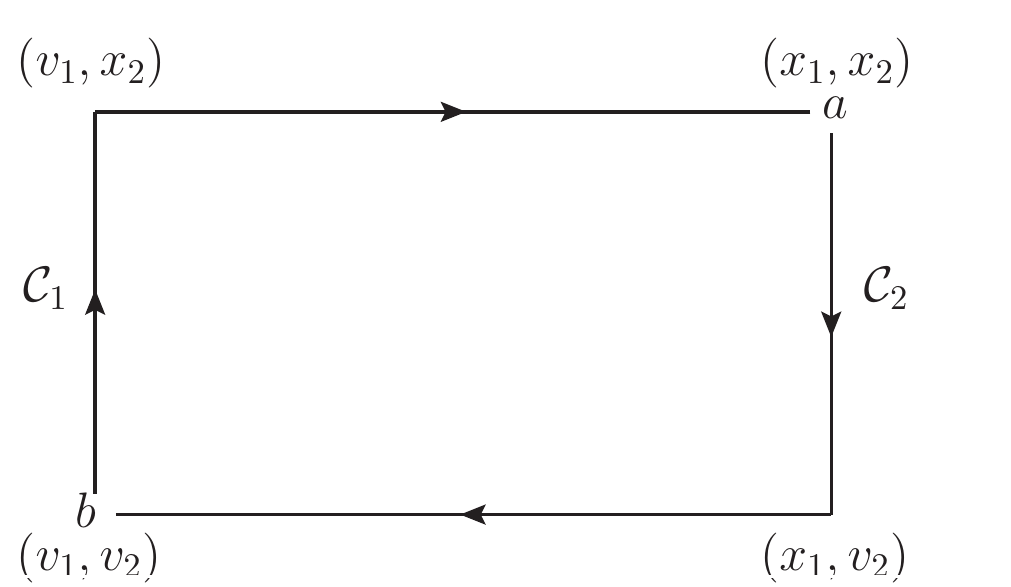}}
\caption{\it Curves $\mathcal{C}_1$ and $\mathcal{C}_2$ used to define $\Phi_{ab}(\vec x,\vec v)$ in Eqs.~(\ref{Phiabe2}) and (\ref{Phi}). \label{fig:curve}}
\end{figure}

The physical results should be independent of the curve $\mathcal{C}$. Nevertheless, for convenience, 
we choose the Wilson line to be symmetric under the combined interchange of color indices and endpoints:
\be
\Phi_{ab}(\mathcal{C};\vec x,\vec v) = \Phi_{ba}(\mathcal{C};\vec v,\vec x)\,.
\ee
For the computations in perturbation theory we need an explicit realization of the Wilson line. We choose the symmetric combination of two paths that go in straight lines (see Fig.~\ref{fig:curve}), so that up to ${\cal O}(e^2)$ the Wilson line reads:
\bE{rCl}
\label{Phiabe2}
\Phi_{ab}(\vec x,\vec v)  &\equiv& \half\left(\Phi_{ab}(\mathcal{C}_1;\vec x,\vec v) +\Phi_{ba}(\mathcal{C}_2;\vec v,\vec x) \right) \\
&=& \de_{ab}-\frac{e}{2}\left(\int_{v_2}^{x_2}\d s_2 A_2(v_1,s_2) + \int_{v_1}^{x_1}\d s_1 A_1(s_1,x_2)\right)_{ab} \nn\\
&&\qquad -\frac{e}{2}\left(\int_{x_2}^{v_2}\d s_2 A_2(x_1,s_2) + \int_{x_1}^{v_1}\d s_1 A_1(s_1,v_2)\right)_{ba} \nn\\
&&\qquad +\frac{(-e)^2}{2}\left(\int_{v_2}^{x_2}\d s_2 A_2(v_1,s_2) \int_{v_2}^{s_2}\d s'_2 A_2(v_1,s'_2) + \int_{v_1}^{x_1}\d s_1 A_1(s_1,x_2) \int_{v_1}^{s_1}\d s'_1 A_1(s'_1,x_2) \right. \nn\\
&&\qquad\qquad\qquad\qquad \left. + \int_{v_1}^{x_1}\d s_1 A_1(s_1,x_2) \int_{v_2}^{x_2}\d s_2 A_2(v_1,s_2) \right)_{ab} \nn\\
&&\qquad +\frac{(-e)^2}{2}\left(\int_{x_2}^{v_2}\d s_2 A_2(x_1,s_2) \int_{x_2}^{s_2}\d s'_2 A_2(x_1,s'_2) + \int_{x_1}^{v_1}\d s_1 A_1(s_1,v_2) \int_{x_1}^{s_1}\d s'_1 A_1(s'_1,v_2) \right. \nn\\
&&\qquad\qquad\qquad\qquad \left. + \int_{x_1}^{v_1}\d s_1 A_1(s_1,v_2) \int_{x_2}^{v_2}\d s_2 A_2(x_1,s_2) \right)_{ba} + {\cal O}(e^3)
\nn
\eE
Note that $A_i^{ab}=-f^{abc}A_i^c$ and $(A_iA_j)^{ab}=f^{adc}f^{dbe}A_i^cA_j^e$. 

It is possible to write $\Phi_{ab}(\vec x,\vec v) $ in a more compact way using the Bars variables \cite{Bars:1978xy}:
\be
 \Phi_{ab}(\vec x,\vec v) ={1\over 2}\left((M_1(\vec x)M_1^{-1}(v_1,x_2)M_2(v_1,x_2)M_2^{-1}(\vec v))^{ab}+(M_2(\vec x)M_2^{-1}(x_1,v_2)M_1(x_1,v_2)M_1^{-1}(\vec v))^{ab}\right)\,, \label{Phi}
\ee
where
\be
M_i(\vec x)=\mathcal{P}e^{-e\int^{\vec x}_{\infty}\d z^iA_i(\vec z)} 
\ee
represents the Wilson line for a straight spatial curve $\mathcal{C}$ with fixed $x^j$ for $j\not=i$. 
This Wilson line can be Taylor expanded in the standard way in terms of (path-ordered) one-dimensional integrals (similarly as 
we have done in Eq.~(\ref{Phiabe2})), or in terms of 
(formal) two dimensional integrals (see, for instance, Ref.~\cite{Freidel:2006qz}):
\begin{IEEEeqnarray}{rCl}
M_i(\vec x) &= 1&-e\int_y G_i(\vec x;\vec y) A_i(\vec y) + e^2\int_{y,z}G_i(\vec x;\vec z)A_i(\vec z)G_i(\vec z;\vec y)A_i(\vec y)+\ldots, \label{M_iofA_i}\\
M_i^{-1}(\vec x) &= 1&+e\int_y G_i(\vec x;\vec y) A_i(\vec y) - e^2\int_{y,z}G_i(\vec x;\vec z)A_i(\vec z)G_i(\vec z;\vec y)A_i(\vec y) \nn\\
&& +  e^2\int_{y,z}G_i(\vec x;\vec z)A_i(\vec z)G_i(\vec x;\vec y)A_i(\vec y)+\ldots\nn\\
\Bigg(&= 1&+e\int_y G_i(\vec x;\vec y) A_i(\vec y) +  e^2\int_{y,z}G_i(\vec x;\vec y)G_i(\vec y;\vec z)A_i(\vec z)A_i(\vec x)+\ldots\Bigg) \,,
\end{IEEEeqnarray}
where
\be
G_1(\vec x;\vec y)\equiv G_1(\vec x-\vec y)=\theta(x_1-y_1)\de(x_2-y_2)\quad \mathrm{and}\quad G_2(\vec x;\vec y)\equiv G_(\vec x-\vec y)=\de(x_1-y_1)\theta(x_2-y_2) \label{G_i}\,.
\ee

Note that $D_iM_i=0$ (no sum over repeated indices). One can also work in the adjoint
\be M_i^{ab}=2\tr(T^aM_iT^bM_i^{-1}) \label{M_iab}
\,,
\ee  and \be D_i^{ab}(\vec y)M_i^{bc}(\vec y) = (\p_i^y\de^{ab}-ef^{abd}A_i^d(\vec y))M_i^{bc}(\vec y)  =0 \label{D_iA_i} 
\,.
\ee 
Starting from this equation
we can compute the functional derivative of this object with respect to $A_j$ and obtain $(\delta M_i)/(\delta A_j)$:
\bE{rCl}
\frac{\de}{\de A_j^e (\vec x)}D_i^{ab}(\vec y)M_i^{bc}(\vec y) &=& -ef^{abe}\de_{ij}\de(\vec y-\vec x)M_i^{bc}(\vec y)  + D_i^{ab}(\vec y) \frac{\de M_i^{bc}(\vec y)}{\de A_j^e (\vec x)}=0 \\
\Longleftrightarrow  \frac{\de M_i^{bc}(\vec y)}{\de A_j^e (\vec x)} &=& e\int_z \left(D_i^{-1}\right)^{ba}_{yz} f^{afe}\de_{ij}\de(\vec z-\vec x)M_i^{fc}(\vec z) \\
&=& e\de_{ij}[M_i(\vec y)G_i(\vec{y}-\vec{x})M_i^{-1}(\vec x)]^{ba} f^{afe} M_i^{fc}(\vec x) \\
&=& e\de_{ij}M_i^{bg}(\vec y)G_i(\vec{y}-\vec{x}) f^{gch} M_i^{eh}(\vec x) \,. 
\eE
In the fundamental representation the derivative of $M_j$ is given by 
\bE{rCl}
 \frac{\de M_j(\vec y)}{\de A_i^a (\vec x)} &=& ie \de_{ij} M_j(\vec y)T^d G_i(\vec y, \vec x)M_i^{ed}(\vec x) \,. \label{M_iOverA_j}
\eE
This can easily be checked by plugging it into the definition of $M_i^{ab}$, Eq.~(\ref{M_iab}).

\medskip

The functional derivative of $A_i$ acting on the Wilson line in Eq.~(\ref{Treg_middleString}) is ill-defined if both the derivative and the Wilson line are defined at the same point. Therefore, we have to regularize it, taking the coincidence limit only after the functional derivative has been applied:
\bE{rl}
\int_{x,v}&\de_\mu(\vec x,\vec v) \left[\frac{\de}{\de A_i^a(\vec x)}\Phi_{ab}(\vec x,\vec v)\right]\frac{\de}{\de A_i^b(\vec v)}  \nn\\
&:= \lim_{\nu\to\infty} \int_{x,v,X}\de_\mu(\vec x,\vec v)\de_\nu(\vec X)\Phi_{ar}(\vec x,\vec x+\vec X) \left[\frac{\de}{\de A_i^r(\vec x+\vec X)}\Phi_{ab}(\vec x,\vec v)\right]\frac{\de}{\de A_i^b(\vec v)} \,. \label{regDeriv}
\eE
This way of regularizing is analogous to the regularizations used in Eq.~(3.24) of Ref.~\cite{Karabali:1997wk} and in Eqs.~(100-101) of Ref.~\cite{Freidel:2006qz}.

Using Eqs.~(\ref{Phi}) and (\ref{M_iOverA_j}) in Eq.~(\ref{regDeriv}) one finds
\be
\int_{x,v}\de_\mu(\vec x,\vec v) \left[\frac{\de}{\de A_i^a(\vec x)}\Phi_{ab}(\vec x,\vec v)\right]\frac{\de}{\de A_i^b(\vec v)}  =0 \,,
\ee
such that the regulated kinetic operator Eq.~(\ref{Treg_middleString}) reduces to
\be
\mathcal{T}_{reg}=-{1\over2}\int_{x,v} \delta_\mu(\vec x, \vec v) \Phi_{ab}(\vec x,\vec v) \frac{\delta}{\delta A_i^a(\vec x)}  \frac{\delta}{\delta A_i^b(\vec v)}\,.
\label{Treg}
\ee
This is shown in Appendix \ref{sec:herm} in detail.

\medskip

Once we have regulated the kinetic operator we turn to the determination of the vacuum wave functional. Realizing that the vacuum wave functional for the kinetic operator $\T$ alone is the identity, one can write the complete wave functional as\footnote{This is possible because the ground state wave function is expected to be real and have zero nodes.}
\be
\Psi=e^{-F}\mathds{1}\,.
\ee
Therefore, instead of solving 
\be
\H\Psi=(\T+\V)\Psi=0
\,,
\ee
one can solve (see, for instance, Ref. \cite{Karabali:1998yq})
\be
\label{HamEq}
\widetilde{\H}\mathds{1}=e^F(\T+\V)e^{-F}\mathds{1}=\left(\T+\V-[\T,F]+\half\left[[\T,F],F\right]\right)
\mathds{1}=0\,,
\ee
since $\T$ contains at most two functional derivatives:
\be
\T=\int_{x}\omega_i^a(\vec x)\frac{\delta}{\delta A_i^a(\vec x)}+\int_{x,y}\Omega_{ij}^{ab}(\vec x,\vec y)\frac{\delta^2}{\delta A_i^a(\vec x) \delta A_j^b(\vec y)} \,,
\ee
where $\omega^a_i(\vec x)=0$ and $\Omega_{ij}^{ab}(\vec x,\vec y)=\de_{ij}\Omega^{ab}(\vec x,\vec y)=-\half\de_{ij}\delta_\mu(\vec x,\vec y)\Phi_{ab}(\vec x,\vec y)$. Using this explicit expression, Eq. (\ref{HamEq}) reads
\be
\V-\int_{x}\omega_i^a(\vec x)\frac{\delta F}{\delta A_i^a(\vec x)}-\int_{x,y}\Omega_{ij}^{ab}(\vec x,\vec y)\frac{\delta^2 F}{\delta A_i^a(\vec x) \delta A_j^b(\vec y)}
+ \int_{x,y}\Omega_{ij}^{ab}(\vec x,\vec y)\frac{\delta F}{\delta A_i^a(\vec x)} \frac{\delta F}{\delta A_j^b(\vec y)} = 0  \label{SE}
\,.
\ee

In order to ensure that we restrict ourselves to gauge invariant states we also have to demand that $\Psi$ satisfies the Gauss law constraint:
\be
\label{Gausslaw}
I^a\Psi =\,
(\vec{D}\cdot\vec{E})^a\Psi =
\left(
\vec \nabla \cdot \frac{\delta }{\delta {\vec A}_a}+ef^{abc}{\vec A}_b \cdot \frac{\delta }{\delta {\vec A}_c}
\right)\Psi=0 
\,.
\ee

In the following we will distinguish between methods (A) and (B), and name their solutions $\Psi_{GL}=e^{-F_{GL}}$ and 
 $\Psi_{GI}=e^{-F_{GI}}$, respectively. The first method consists in directly solving Eqs.~(\ref{SE}) and (\ref{Gausslaw}), and will be addressed in the next section. The second method consists in rewriting Eq.~(\ref{SE}) in terms of the gauge invariant variables proposed in Refs. \cite{Karabali:1995ps,Karabali:1996je,Karabali:1996iu, Karabali:1997wk,Karabali:1998yq}. It will be addressed in Sec.~\ref{sec:KKN}. In both cases we will Taylor expand $F$ in powers of the coupling constant $e$, and solve the resulting equations iteratively. A detailed explanation of both computations can be found in Ref.~\cite{Krug:2013yq}. In this paper the main focus will be on the novel aspects resulting from the careful introduction of the regularization. 

\section{Determination of $\Psi_{GL}[{\vec A}]$}
\label{sec:Hatfield} 
We expand $F_{GL}=F_{GL}^{(0)}+eF_{GL}^{(1)}+e^2F_{GL}^{(2)}+{\cal O}(e^3)$ and 
\bE{l}
\Omega^{ab}(\vec x,\vec y)=-{1\over2}\delta_\mu(\vec x,\vec y)\Phi_{ab}(\vec x,\vec y)=-{1\over2}\delta_\mu(\vec x,\vec y)\left(\Phi^{(0)}_{ab}(\vec x,\vec y)+e\Phi^{(1)}_{ab}(\vec x,\vec y)+e^2\Phi^{(2)}_{ab}(\vec x,\vec y)+{\cal O}(e^3)\right) \qquad
\eE
in powers of $e$, the coupling constant. Considering the contributions order by order in $e$ yields the following equations:

\medskip

At ${\cal O}(e^0)$ we have
\be
\label{eqe0}
\V|_{{\cal O}(e^0)} -{1\over2}\int_{x,y}\delta_\mu(\vec x,\vec y)\delta_{ab} \left(- \frac{\delta^2 F_{GL}^{(0)} }{\delta A_i^a(\vec x) \delta A_i^b(\vec y)}+\frac{\delta F_{GL}^{(0)}}{\delta A_i^a(\vec x)} \frac{\delta F_{GL}^{(0)}}{\delta A_i^b(\vec y)} \right)= 0\,.
\ee
It is easier to solve it in momentum space using
\be
{\vec A}(\vec x)=\int_\slashed{k} {\vec A}(\vec k)e^{i{\vec k}\cdot{\vec x}}
\,,
\qquad
\frac{\delta}{\delta \vec{A}^a(\vec{x})}
=\int_\slashed{k} 
\frac{\delta}{\delta \vec{A}^a(\vec{k})}e^{-i{\vec k}\cdot{\vec x}}
\,.\label{FT}
\ee
For Eq. (\ref{eqe0}) we can take the $\mu\to\infty$ limit, reducing it to the standard unregulated free field equation, the solution of which is known and reads
\bea
F_{GL}^{(0)}[{\vec A}] &=& \frac{1}{2}\int_\slashed{k}\frac{1}{|\vec k|} (\vec{k}\times\vec{A}^a(\vec{k})) (\vec{k}\times\vec{A}^a(-\vec{k})) \label{FGL0m} \\
&=& \frac{1}{4\pi}\int_{x,y}\frac{1}{|\vec{x}-\vec{y}|} (\vec{\nabla}\times\vec{A}^a(\vec{x}))(\vec{\nabla}\times\vec{A}^a(\vec{y})) 
\,.\label{FGL0p} 
\eea
 
\medskip

At ${\cal O}(e^1)$ we have
\bE{l}
\V|_{{\cal O}(e^1)} +{1\over2}\int_{x,y}\delta_\mu(\vec x,\vec y)\delta_{ab}
\left(
\frac{\delta^2 F_{GL}^{(1)}}{\delta A_i^a(\vec x) \delta A_i^b(\vec y)} 
- 2  \frac{\delta F_{GL}^{(0)}}{\delta A_i^a(\vec x)} \frac{\delta F_{GL}^{(1)}}{\delta A_i^b(\vec y)}
\right) 
\nn\\
\quad -{1\over2} \int_{x,y}\delta_\mu(\vec x,\vec y)\Phi^{(1)}_{ab}(\vec x,\vec y) 
\left(
\frac{\delta F_{GL}^{(0)}}{\delta A_i^a(\vec x)} \frac{\delta F_{GL}^{(0)}}{\delta A_i^b(\vec y)} 
- \frac{\delta^2 F_{GL}^{(0)}}{\delta A_i^a(\vec x) \delta A_i^b(\vec y)}
\right) = 0\,.
\eE
Both terms proportional to $\Phi^{(1)}_{ab}(\vec x,\vec y)$ vanish (the second because of contraction of color indices, for the first see App.~\ref{subsec:append4}). For the remaining terms we can take the limit $\mu \rightarrow \infty$. Therefore, this equation also reduces to the unregularized \SE, which we already took care of in Ref.~\cite{Krug:2013yq}. It is solved by
\begin{IEEEeqnarray}{rCl}
\label{FGL1}
F_{GL}^{(1)}[{\vec A}] &=&  i f^{abc} \int_{\slashed{k_1},\slashed{k_2},\slashed{k_3}}\slashed{\delta}\left(\sum_{i=1}^3 \vec{k}_i\right) \Bigg\{ \frac{1}{2(\sum_i^3|\vec{k}_i|)} (\vec{k}_1\times\vec{A}^a(\vec{k}_1)) (\vec{A}^b(\vec{k}_2)\times\vec{A}^c(\vec{k}_3))\nn\\
&& -\frac{1}{(\sum_i^3|\vec{k}_i|)|\vec{k}_1||\vec{k}_3|} (\vec{k}_1\cdot\vec{A}^a(\vec{k}_1)) (\vec{k}_3\times\vec{A}^b(\vec{k}_2)) (\vec{k}_3\times\vec{A}^c(\vec{k}_3)) \Bigg\} 
\,.
\end{IEEEeqnarray}

\medskip

At ${\cal O}(e^2)$ we determine $F_{GL}^{(2)}$. $F^{(2)}_{GL}$ can have contributions with four, two and zero fields: $F^{(2)}_{GL}=F^{(2,4)}_{GL}+F^{(2,2)}_{GL}+F^{(2,0)}_{GL}$. There is no need to compute $F^{(2,0)}_{GL}$, as it only changes the normalization of the state, which we do not fix, or alternatively can be absorbed in a redefinition of the ground-state energy. $F^{(2,4)}_{GL}$ is determined by the following equation:
\bE{l}
 \V|_{{\cal O}(e^2)}   -{1\over2} \int_{x,y}\delta_\mu(\vec x,\vec y)\delta_{ab}\left( \frac{\delta F_{GL}^{(1)}}{\delta A_i^a(\vec x)} \frac{\delta F_{GL}^{(1)}}{\delta A_i^b(\vec y)}  + 2\frac{\delta F_{GL}^{(0)}}{\delta A_i^a(\vec x)} \frac{\delta F_{GL}^{(2,4)}}{\delta A_i^b(\vec y)} \right)  \nn\\
\quad  -{1\over2}\int_{x,y}\delta_\mu(\vec x,\vec y)\left(\Phi^{(2)}_{ab}(\vec x,\vec y) \frac{\delta F_{GL}^{(0)}}{\delta A_i^a(\vec x)} \frac{\delta F^{(0)}}{\delta A_i^b(\vec y)} + 2 \Phi^{(1)}_{ab}(\vec x,\vec y) \frac{\delta F_{GL}^{(0)}}{\delta A_i^a(\vec x)} \frac{\delta F_{GL}^{(1)}}{\delta A_i^b(\vec y)} \right) = 0 \,,
\label{regF24eq}
\eE
The two terms in the second line vanish (see App.~\ref{subsec:append5}). For the leftover we can take the $\mu \rightarrow \infty$ limit. Eq.~(\ref{regF24eq}) then reduces to its unregularized version, which was solved in Ref.~\cite{Krug:2013yq}. We quote it here for completeness:
\begin{IEEEeqnarray}{l}
\label{F24GL}
F_{GL}^{(2,4)}[\vec A]=f^{abc}f^{cde}\int_{\slashed{k_1},\slashed{k_2},\slashed{q_1},\slashed{q_2}}
\slashed{\delta}\left(\sum_i (\vec{k}_i+\vec{q}_i)\right) \frac{1}{|\vec{k}_1|+|\vec{k}_2|+|\vec{q}_1|+|\vec{q}_2|} \Bigg\{  \nn\\
\quad \frac{1}{2(|\vec{k}_1|+|\vec{k}_2|+|\vec{k}_1+\vec{k}_2|)(|\vec{q}_1|+|\vec{q}_2|+|\vec{q}_1+\vec{q}_2|)} \Bigg\{\left(\vec{A}^d(\vec{q}_1)\times\vec{A}^e(\vec{q}_2)\right)
\Bigg[ -\frac{1}{4}|\vec{k}_1+\vec{k}_2|^2 \vec{A}^a(\vec{k}_1)\times\vec{A}^b(\vec{k}_2) \nn\\
\qquad\qquad +\frac{|\vec{k}_1+\vec{k}_2|}{|\vec{k}_2|}(\vec{k}_1+\vec{k}_2)\times\vec{A}^a(\vec{k}_1) (\vec{k}_2\cdot\vec{A}^b(\vec{k}_2)) +\frac{(\vec{k}_1+\vec{k}_2)\cdot\vec{k}_2}{|\vec{k}_1||\vec{k}_2|} (\vec{k}_1\cdot\vec{A}^a(\vec{k}_1)) (\vec{k}_2\times\vec{A}^b(\vec{k}_2)) \nn\\
\qquad\qquad +(\vec{k}_1\times\vec{A}^a(\vec{k}_1)) (\vec{k}_1+\vec{k}_2)\cdot\vec{A}^b(\vec{k}_2) \Bigg] \nn\\
\qquad +  (\vec{k}_1\times\vec{A}^a(\vec{k}_1)) (\vec{q}_1\times\vec{A}^d(\vec{q}_1)) \left(\vec{A}^b(\vec{k}_2)\cdot\vec{A}^e(\vec{q}_2)\right)\nn\\
\qquad + \frac{1}{|\vec{k}_1||\vec{k}_2|}\left[2\vec{k}_2\cdot\vec{A}^e(\vec{q}_2)-\frac{\vec{q}_1\cdot\vec{k}_2}{|\vec{q}_1||\vec{k}_2|}\vec{q}_2\cdot\vec{A}^e(\vec{q}_2)\right] (\vec{k}_1\cdot\vec{A}^a(\vec{k}_1)) (\vec{k}_2\times\vec{A}^b(\vec{k}_2)) (\vec{q}_1\times\vec{A}^d(\vec{q}_1)) \nn\\
\qquad + \frac{1}{|\vec{k}_1|} (\vec{k}_1\cdot\vec{A}^a(\vec{k}_1)) (\vec{k}_1+\vec{k}_2) \times\vec{A}^b(\vec{k}_2) \Bigg[  \frac{1}{|\vec{q}_2|} (\vec{q}_1+\vec{q}_2)\times\vec{A}^d(\vec{q}_1) (\vec{q}_2 \cdot\vec{A}^e(\vec{q}_2)) \nn\\
\qquad\qquad + \frac{2}{|\vec{q}_1+\vec{q}_2|} (\vec{q}_1\times\vec{A}^d(\vec{q}_1)) (\vec{q}_1+\vec{q}_2) \cdot\vec{A}^e(\vec{q}_2) \Bigg] \nn\\
\qquad -\frac{2(\vec{q}_1+\vec{q}_2)\cdot\vec{q}_1}{|\vec{k}_1+\vec{k}_2||\vec{k}_1| |\vec{q}_1||\vec{q}_2|}  (\vec{k}_1\cdot\vec{A}^a(\vec{k}_1)) (\vec{k}_1+\vec{k}_2) \times\vec{A}^b(\vec{k}_2) (\vec{q}_1\times\vec{A}^d(\vec{q}_1)) (\vec{q}_2 \cdot\vec{A}^e(\vec{q}_2)) \nn\\
\qquad +\frac{2\vec{k}_1\times\vec{k}_2}{|\vec{k}_1||\vec{k}_2| |\vec{q}_1+\vec{q}_2| |\vec{q}_2|}  (\vec{k}_1\cdot\vec{A}^a(\vec{k}_1)) (\vec{k}_2 \times\vec{A}^b(\vec{k}_2)) (\vec{q}_2\times\vec{A}^d(\vec{q}_1)) (\vec{q}_2 \times\vec{A}^e(\vec{q}_2)) \nn\\
\qquad +\frac{2}{|\vec{q}_1+\vec{q}_2||\vec{q}_2|}  (\vec{k}_1\times\vec{A}^a(\vec{k}_1)) (\vec{k}_1+\vec{k}_2) \times\vec{A}^b(\vec{k}_2) (\vec{q}_2\times\vec{A}^d(\vec{q}_1)) (\vec{q}_2 \times\vec{A}^e(\vec{q}_2)) \nn\\
\qquad -\frac{1}{|\vec{k}_2||\vec{q}_2|}  (\vec{k}_2\times\vec{A}^a(\vec{k}_1)) (\vec{k}_2 \times\vec{A}^b(\vec{k}_2)) (\vec{q}_2\times\vec{A}^d(\vec{q}_1)) (\vec{q}_2 \times\vec{A}^e(\vec{q}_2))
\Bigg\} \nn\\
\quad +\frac{1}{8} \left(\vec{A}^a(\vec{k}_1)\times\vec{A}^b(\vec{k}_2)\right) \left(\vec{A}^d(\vec{q}_1)\times\vec{A}^e(\vec{q}_2)\right) \nn\\
\quad + \frac{1}{|\vec{k}_1|(|\vec{q}_1|+|\vec{q}_2|+|\vec{q}_1+\vec{q}_2|)}  (\vec{k}_1\cdot\vec{A}^a(\vec{k}_1)) \Bigg\{ \frac{1}{2}  (\vec{k}_1+\vec{k}_2) \times\vec{A}^b(\vec{k}_2)  \left(\vec{A}^d(\vec{q}_1)\times\vec{A}^e(\vec{q}_2)\right) \nn\\
\quad\quad 
-  (\vec{q}_1 \times\vec{A}^d(\vec{q}_1)  \left(\vec{A}^b(\vec{k}_2)\times\vec{A}^e(\vec{q}_2)\right) -  \frac{1}{|\vec{q}_1+\vec{q}_2||\vec{q}_2|} (\vec{k}_1+\vec{k}_2) \times\vec{A}^b(\vec{k}_2) (\vec{q}_1+\vec{q}_2)\times\vec{A}^d(\vec{q}_1) (\vec{q}_2 \cdot\vec{A}^e(\vec{q}_2)) \nn\\
\quad \quad +  \frac{1}{|\vec{q}_1||\vec{q}_2|} (\vec{q}_2 \times\vec{A}^b(\vec{k}_2)) (\vec{q}_1\cdot\vec{A}^d(\vec{q}_1)) (\vec{q}_2 \times\vec{A}^e(\vec{q}_2)) \nn\\
\quad \quad  -  \frac{1}{|\vec{q}_1+\vec{q}_2||\vec{q}_2|} (\vec{k}_1+\vec{k}_2) \cdot\vec{A}^b(\vec{k}_2) (\vec{q}_2\times\vec{A}^d(\vec{q}_1)) (\vec{q}_2 \times\vec{A}^e(\vec{q}_2)) 
\Bigg\}
\Bigg\}
\,. 
\end{IEEEeqnarray}

So far the regularization of the kinetic term has not produced any modification to the results obtained in Ref.~\cite{Krug:2013yq}. This could have been expected. If we have to make an analogy 
of this computation to the standard diagrammatic approach, the computations above would correspond to tree-level-like diagrams, for which one  
can take the cutoff to infinity. It is only when one has internal loops, where the momentum can run to infinity, when regularization effects become important. In our approach those effects are hidden in $F^{(2,2)}_{GL}$, where we have a similar effect to the contraction of two fields. We compute this term in the 
next subsection.

\subsection{$F^{(2,2)}_{GL}$}
$F^{(2,2)}_{GL}$ is determined by the following equation:
\bE{l}
\int_{x,y}\delta_\mu(\vec x,\vec y)\Bigg(\delta_{ab}\frac{\delta^2 F_{GL}^{(2,4)} }{\delta A_i^a(\vec x) \delta A_i^b(\vec y)} + \Phi^{(1)}_{ab}(\vec x,\vec y)\frac{\delta^2 F_{GL}^{(1)} }{\delta A_i^a(\vec x) \delta A_i^b(\vec y)}+ \Phi^{(2)}_{ab}(\vec x,\vec y)\frac{\delta^2 F_{GL}^{(0)} }{\delta A_i^a(\vec x) \delta A_i^b(\vec y)} \nn\\
\qquad - 2 \delta_{ab} \frac{\delta F_{GL}^{(0)}}{\delta A_i^a(\vec x)} \frac{\delta F_{GL}^{(2,2)}}{\delta A_i^b(\vec y)} \Bigg) = 0 \,.
\label{regF22eq}
\eE
In order to solve this equation it is convenient to rewrite it in momentum space. Then, the last term of Eq.~(\ref{regF22eq}) reads
\bea
&&- 2 \int_{x,y} \delta_\mu(\vec x,\vec y) \Phi^{(0)}_{ab}(\vec x,\vec y) \frac{\delta F_{GL}^{(0)}}{\delta A_i^a(\vec x)} \frac{\delta F_{GL}^{(2,2)}}{\delta A_i^b(\vec y)}  \nn\\
&=&-2 \int_{\slashed{p}}\delta_\mu(\vec p)\delta^{ab} \frac{1}{|{\vec p}|} (\vec{p}\times\vec{A}^a(\vec{p}))  \left(\vec{p}\times \frac{\delta F_{GL}^{(2,2)}[\vec A]}{\delta \vec{A}^b(\vec{p})}\right) 
\nn\\
&=&-2 \int_{\slashed{p}}\delta_\mu(\vec p)\frac{1}{|{\vec p}|} 
\left\{
{\vec p}^2\left(\vec{A}^a(\vec{p})\cdot \frac{\delta F_{GL}^{(2,2)}[\vec A]}{\delta \vec{A}^a(\vec{p})}\right) 
-
\left({\vec p}\cdot {\vec A}^a({\vec p})\right)
\left(\vec{p}\cdot \frac{\delta F_{GL}^{(2,2)}[\vec A]}{\delta \vec{A}^a(\vec{p})}\right) 
\right\}
\,, \label{edd}
\eea
where $\de_\mu(\vec p)=e^{-\frac{\vec{p}^2}{4\mu^2}}$ is the Fourier transform of $\delta_\mu(\vec x,\vec y)$ and we used $\e_{ij}\e_{kl}=\de_{ik}\de_{jl}-\de_{jk}\de_{il}$.\\
The Gauss law implies that the second term on the right-hand-side of the last equality of Eq.~(\ref{edd}) vanishes, so Eq.~(\ref{regF22eq}) can be rewritten as 
\bE{rCl}
\label{F22mom}
2 \int_{\slashed{p}}\delta_\mu(\vec p)|{\vec p}| \left(\vec{A}^a(\vec{p})\cdot \frac{\delta F_{GL}^{(2,2)}[\vec A]}{\delta \vec{A}^a(\vec{p})}\right) 
 &=& \int_{x,y}\int_{\slashed{p},\slashed{q}}e^{-i{\vec p}\cdot{\vec x}}e^{-i{\vec q}\cdot{\vec y}}\delta_\mu(\vec x,\vec y)\Bigg(\de^{ab}\frac{\delta^2 F_{GL}^{(2,4)} }{\delta A_i^a(\vec p) \delta A_i^b(\vec q)} \nn\\
&&\quad + \Phi^{(1)}_{ab}(\vec x,\vec y)\frac{\delta^2 F_{GL}^{(1)} }{\delta A_i^a(\vec p) \delta A_i^b(\vec q)}+ \Phi^{(2)}_{ab}(\vec x,\vec y)\frac{\delta^2 F_{GL}^{(0)} }{\delta A_i^a(\vec p) \delta A_i^b(\vec q)} \Bigg) \,. \qquad
\eE
Before going on we need to compute the right-hand-side of this equation (which again is better handled in momentum space). The first term corresponds to the regulated version of the term that already appeared in Ref.~\cite{Krug:2013yq}. 
As we can see in Eq.~(\ref{F24GL}), the explicit expression of $F^{(2,4)}_{GL}[\vec A]$ is very lengthy and complicated. 
This made impossible a direct brute force computation of $\frac{\delta^2 F_{GL}^{(2,4)} }{\delta A_i^a(\vec x) \delta A_i^b(\vec y)}$. The strategy we followed 
instead was to rewrite $F^{(2,4)}_{GL}[\vec A]$ in terms of $J$ and $\theta=\frac{1}{\bar\p}\bar A+ \mathcal{O}(e)$ (see next section for notation), which allows for a cleaner arrangement of the terms (see Eq.~(82) of \cite{Krug:2013yq}), in particular between gauge invariant and gauge dependent terms. Proceeding analogously to this reference and using (see Eq.~(78) of \cite{Krug:2013yq})
\begin{IEEEeqnarray}{rCl}
\int_p\frac{\delta^2}{\delta A^a_i(-\vec{p})\delta A^a_i(\vec{p})} 
&=& 4 \int_p\frac{p}{\bar{p}}\frac{\delta^2}{\delta J^a(-\vec{p})\delta J^a(\vec{p})} +2\int_p \bar{p} \frac{\delta^2}{\delta \theta^a(-\vec{p})\delta J^a(\vec{p})} +\mathcal{O}(e)\,,
\end{IEEEeqnarray}
where the Fourier transformation for $J^a$ and $\theta^a$ is defined analogously to Eq.~(\ref{FT}), 
we obtain
\bea
\label{F22Jtheta}
&& 
\int_{x,y}\int_{\slashed{p},\slashed{q}}e^{-i{\vec p}\cdot{\vec x}}e^{-i{\vec q}\cdot{\vec y}}\delta_\mu(\vec x,\vec y)\de^{ab}\frac{\delta^2 F_{GL}^{(2,4)} }{\delta A_i^a({\vec p}) \delta A_i^b({\vec q})} 
 = 
 \\
 &&
 4C_A \int_{\slashed{p},\slashed{k}} e^{-\frac{\vec{p}^2}{4\mu^2}} \Bigg\{  \left(-\frac{1}{32} \frac{1}{\bar{p}}  g^{(3)}(\vec{k},\vec{p},-\vec{k}-\vec{p}) - \frac{1}{64} \frac{p}{\bar{p}}  g^{(4)}(\vec{p},\vec{k};-\vec{p},-\vec{k}) \right) J^{a}(\vec{k})J^{a}(-\vec{k}) \nn\\
&& + \frac{1}{4}\Bigg(\frac{1}{4}\left(2\frac{p}{\bar{p}}+\frac{k}{\bar{k}+\bar{p}}-\frac{p\bar{k}}{\bar{p}(\bar{k}+\bar{p})}\right) g^{(3)}(\vec{p},\vec{k},-\vec{p}-\vec{k}) \nn\\
 &&\qquad  -2 \frac{1}{\bar{p}} \frac{(\bar{k}+\bar{p})^2}{|\vec{k}+\vec{p}|} +2 \frac{1}{\bar{p}} \frac{\bar{k}^2}{|\vec{k}|}  - \frac{\bar{k}-\bar{p}}{\bar{p}(\bar{k}+\bar{p})} \frac{\bar{k}^2}{|\vec{k}|} + \frac{\bar{k}-\bar{p}}{\bar{k}+\bar{p}} \frac{\bar{p}}{|\vec{p}|}\Bigg) J^{a}(\vec{k})\theta^{a}(-\vec{k}) \nn\\
 && \hspace{-1cm}+ \left( \frac{p}{\bar{p}} \left( \frac{(\bar{p}+\bar{k})^2}{|\vec{p}+\vec{k}|} -  \frac{\bar{p}^2}{|\vec{p}|} \right) - \frac{p}{\bar{p}}\bar{k} \left(\frac{\bar{p}+\bar{k}}{|\vec{p}+\vec{k}|} - \frac{\bar{p}}{|\vec{p}|} \right) + k \left(\frac{\bar{p}+\bar{k}}{|\vec{p}+\vec{k}|} - \frac{\bar{p}}{|\vec{p}|} \right) \right) \theta^{a}(\vec{k}) \theta^{a}(-\vec{k}) \Bigg\} \nn
\,. 
\eea
This expression has an internal loop for the momentum $\vec p$, the integral of which is regulated by $\delta_{\mu}({\vec p})$. 
If we naively take the limit $\mu \rightarrow \infty$ and do formal manipulations (momentum shifts) of the integrals, we find the result obtained 
in Ref.~\cite{Krug:2013yq}:
\be
\label{J2I4}
-N\frac{C_A}{\pi}\int_\slashed{k}\frac{\bar{k}^2}{|\vec{k}|^2}J^{a}(\vec{k})J^{a}(-\vec{k}) =-N\frac{C_A}{\pi}\int_\slashed{k}\frac{1}{|\vec{k}|^2} (\vec{k}\times\vec{A}^a(\vec{k})) (\vec{k}\times\vec{A}^a(-\vec{k})) \,,
\ee
\bE{rCl}
\nn
N &=& \frac{|\vec{k}|}{{\bar k}^2}\left(\int \frac{d^2 p}{32\pi}\ \frac{1}{\bar p}\ g^{(3)}(\vec{k},\vec{p},-\vec{p}-\vec{k})\ +\ \int \frac{d^2 p}{64\pi}\ \frac{p}{\bar p}\ g^{(4)}(\vec{k},\vec{p};-\vec{k},-\vec{p})\right) \label{N} \\
&=& 0.025999\,(8\pi) \,,
\eE
whereas the terms proportional to $J \theta$ and $\theta^2$ vanish. Yet, this is not the whole story. 
The internal momentum of the loop is characterized by two scales: $|{\vec p}| \sim \mu$ and $|{\vec p}| \sim |{\vec k}|$, and taking the limit $\mu \rightarrow \infty$ before integration neglects contributions from the $|{\vec p}| \sim \mu$ region. Things change once the regularization is taken into account, as the high energy modes $|{\vec p}| \sim \mu$ are now also included in the computation. 
The loop result of the $J^2$ term is not modified by the introduction of the regularization, since the contribution due to $|{\vec p}| \sim \mu$ is subleading. Therefore, Eq.~(\ref{J2I4}) remains unchanged. 
Things are different, however, for the $J\theta$ and $\theta^2$ term. The $\theta^2$ term can be simplified to the following expression
 \begin{IEEEeqnarray}{rCl}
4 C_A\int_{\slashed{p},\slashed{k}} e^{-\frac{\vec{p}^2}{4\mu^2}}  \left( \left(  \frac{p(\bar{p}+\bar{k})}{|\vec{p}+\vec{k}|} -  \frac{1}{4}|\vec{p}|  \right) + k \left(\frac{\bar{p}+\bar{k}}{|\vec{p}+\vec{k}|} \right)  + \frac{\bar{k}p-k\bar{p}}{|\vec{p}|}\right) \theta^{a}(\vec{k}) \theta^{a}(-\vec{k}) 
\,.
\end{IEEEeqnarray}
The last term vanishes under $\vec p\to-\vec p$ and the first and the third can be combined to yield (note that the integral is dominated by $|{\vec p}| \sim \mu$ and that the $|{\vec p}| \sim |{\vec k}|$ region gives subleading contributions) 
\begin{IEEEeqnarray}{rCl}
\label{theta2}
C_A  \int_{\slashed{p},\slashed{k}} e^{-\frac{\vec{p}^2}{4\mu^2}}  \left(|\vec{p}+\vec{k}| -|\vec{p}|  \right)  \theta^{a}(\vec{k}) \theta^{a}(-\vec{k}) 
=
\int_{ \slashed{k}}\frac{C_A\mu}{8\sqrt{\pi}}  |\vec{k}|^2  \theta^{a}(\vec{k}) \theta^{a}(-\vec{k})+{\cal O}(1/\mu)\,.
\end{IEEEeqnarray}
We can deal with the $J\theta$ term of Eq.~(\ref{F22Jtheta}) in a very similar way (though with lengthier expressions). 
As before, the integral is dominated by the $|{\vec p}| \sim \mu$ region, whereas the $|{\vec p}| \sim |{\vec k}|$ region of momentum gives a subleading contribution\footnote{Actually this sort of statements are not true in general, as finite momentum shifts in the integrals may produce corrections from the $|{\vec p}| \sim |{\vec k}|$ region. Such shifts do not change the leading order contribution, which in our case is of ${\cal O}(\mu)$, but may change the individual ${\cal O}(\mu^0)$ contributions due to the $|{\vec p}| \sim |{\vec k}|$ and $|{\vec p}| \sim \mu$ regions (but in such a way that the total sum remains the same), which is the precision we seek. Therefore, such statements should be understood for a specific routing of momenta.}. Using
\bE{rCl}
{1\over2}\left(  J^{a}(\vec{k})\theta^{a}(-\vec{k}) -  J^{a}(-\vec{k})\theta^{a}(\vec{k}) \right) &=& -\frac{1}{2\bar{k}}\vec{A}^a(\vec{k})\cdot\vec{A}^a(-\vec{k})+2k\;\theta^a(\vec{k})\theta^a(-\vec{k}) +O(e)\,,
\eE
we rewrite the result in terms of $\vec{A}$ and $\theta$, and obtain
\begin{IEEEeqnarray}{rCl}
\label{jtheta}
 -\frac{C_A}{8\sqrt{\pi}} \mu  \int_{\slashed{k}}&    \Bigg(-\vec{A}^a(\vec{k})\cdot\vec{A}^a(-\vec{k})+ |\vec{k}|^2\, \theta^a(\vec{k})\theta^a(-\vec{k}) \Bigg) \,.
\end{IEEEeqnarray}
The bilinear terms in $\theta$ in Eqs.~(\ref{theta2}) and (\ref{jtheta}) cancel each other. Therefore, summing the contributions from Eqs.~(\ref{J2I4}), (\ref{theta2}) and (\ref{jtheta}) we obtain 
\bea
\label{F2I}
&&
\int_{x,y}\int_{\slashed{p},\slashed{q}}e^{-i{\vec p}\cdot{\vec x}}e^{-i{\vec q}\cdot{\vec y}}\delta_\mu(\vec x,\vec y)\de^{ab}\frac{\delta^2 F_{GL}^{(2,4)} }{\delta A_i^a({\vec p}) \delta A_i^b(\vec q)} 
 = 
 \\
 \nn
 &&
-N\frac{C_A}{\pi}\int_\slashed{k}\frac{1}{|\vec{k}|} (\vec{k}\times\vec{A}^a(\vec{k})) (\vec{k}\times\vec{A}^a(-\vec{k})) +\frac{C_A}{8\sqrt{\pi}}\mu \int_{\slashed{k}}  \vec{A}^a(\vec k)\cdot\vec{A}^a(-\vec k) +O\left(\mu^{-1}\right) 
\,.
\eea

Analogously, we compute the second term of the right-hand-side of Eq.~(\ref{F22mom}):
\bE{l}
\label{F2II}
\int_{x,y}\int_{\slashed{p},\slashed{q}}e^{-i{\vec p}\cdot{\vec x}}e^{-i{\vec q}\cdot{\vec y}}\delta_\mu(\vec x,\vec y) \Phi^{(1)}_{ab}(\vec x,\vec y)\frac{\delta^2 F_{GL}^{(1)} }{\delta A_i^a(\vec p) \delta A_i^b(\vec q)}
\nn\\
 = {1\over 2} f^{abd}  \int_{u,v,y}\int_{\slashed{k},\slashed{q},\slashed{p}}\delta_\mu(\vec u,\vec v)   \Bigg\{(G_1(\vec u;\vec y)-G_1(v_1,u_2;\vec y)+G_1(u_1,v_2;\vec y)-G_1(\vec v;\vec y)) A_1^d(\vec y)  \nn\\
\qquad  + (G_2(v_1,u_2;\vec y)-G_2(\vec v;\vec y)+G_2(\vec u;\vec y)-G_2(u_1,v_2;\vec y)) A_2^d(\vec y) \Bigg\} \nn\\
\quad 
\times
 i f^{abc} e^{-i\vec{p}\cdot\vec{u}} e^{-i\vec{q}\cdot\vec{v}} \frac{\slashed{\delta}\left(\vec{k}+\vec{p}+\vec{q}\right)}{|\vec{k}|+|\vec{p}|+|\vec{q}|} \Bigg\{  (\vec{q}-\vec{p})\cdot  \vec{A}^c(\vec{k})   -\frac{|\vec{q}|-|\vec{p}|}{|\vec{k}|} \vec{k}\cdot\vec{A}^c(\vec{k}) \nn\\
\qquad + \frac{1}{|\vec{q}||\vec{k}|}  (\vec{k}\times\vec{q}) (\vec{k}\times\vec{A}^c(\vec{k})) + \frac{1}{|\vec{q}||\vec{p}|} (\vec{p}\times\vec{q}) (\vec{k}\times\vec{A}^c(\vec{k}))  -\frac{1}{|\vec{p}||\vec{k}|}   (\vec{k}\times\vec{p})  (\vec{k}\times\vec{A}^c(\vec{k}))   \Bigg\} 
\nn
\quad\\
 = {C_A\over 2}\int_{\slashed{p},\slashed{q}} \Bigg\{\left(e^{-\frac{(p_1-q_1)^2}{4 \mu ^2}}-e^{-\frac{q_1^2}{4\mu^2}} \right)\left(e^{-\frac{(p_2-q_2)^2}{4 \mu ^2}}+e^{-\frac{q_2^2}{4\mu^2}} \right) \frac{1}{p_1} A_1^c(\vec p)  \nn\\
\qquad  + \left(e^{-\frac{(p_2-q_2)^2}{4 \mu ^2}}-e^{-\frac{q_2^2}{4\mu^2}} \right)\left(e^{-\frac{(p_1-q_1)^2}{4 \mu ^2}}+e^{-\frac{q_1^2}{4\mu^2}} \right) \frac{1}{p_2} A_2^c(\vec p)  \Bigg\}  \nn\\
\quad  
\times
\frac{1}{|\vec{q}|+|\vec{p}|+|\vec{q}-\vec{p}|} \Bigg\{  (\vec{p}-2\vec{q})\cdot  \vec{A}^c(-\vec{p})   -\frac{|\vec{q}-\vec{p}|-|\vec{q}|}{|\vec{p}|} (-\vec{p})\cdot\vec{A}^c(-\vec{p}) \nn\\
\qquad + \Bigg(\frac{ (-\vec{q}\times\vec{p})}{|\vec{p}||\vec{q}-\vec{p}|}  -\frac{(-\vec{p}\times\vec{q})}{|\vec{q}||\vec{p}|}  + \frac{(\vec{q}\times\vec{p})}{|\vec{q}||\vec{q}-\vec{p}|}   \Bigg) (-\vec{p})\times\vec{A}^c(-\vec{p})   \Bigg\} 
\nn
\\
 =- {C_A\over 4\sqrt{\pi}}\mu \int_{\slashed{p}}  \vec{A}^c(-p)\cdot\vec{A}^c(p) - {C_A\over 8 \pi} \int_{\slashed{p}}   \frac{1}{|\vec{p}|}  (\vec{p}\times\vec{A}^c(-\vec{p})) (\vec{p}\times\vec{A}^c(\vec{p}))   + O\left(\mu^{-1}\right)
\,.
\eE
Finally, the third term of the right-hand-side of Eq.~(\ref{F22mom}) reads
\bE{rCl}
\label{F2III}
&&\int_{x,y}\int_{\slashed{p},\slashed{q}}e^{-i{\vec p}\cdot{\vec x}}e^{-i{\vec q}\cdot{\vec y}}\delta_\mu(\vec x,\vec y) \Phi^{(2)}_{ab}(\vec x,\vec y)\frac{\delta^2 F_{GL}^{(0)} }{\delta A_i^a(\vec p) \delta A_i^b(\vec q)}
\nn\\
 &=& 2\frac{1}{4\pi}\int_{u.vx,w} \delta_\mu(\vec u,\vec v) \frac{1}{|\vec{x}-\vec{w}|} \p_{x_i}\delta(\vec x-\vec u) \p_{w_i}\delta(\vec w-\vec v) \delta^{ab} \nn\\
 && \Bigg\{  {1\over 2}f^{adc}f^{dbe}\int_{y,z}\Big( (G_1(\vec u;\vec z)-G_1(v_1,u_2;\vec z))(G_1(\vec z;\vec y)-G_1(v_1,u_2;\vec y))\nn\\
&& \qquad + (G_1(u_1,v_2;\vec z)-G_1(\vec v;\vec z))(G_1(\vec z;\vec y)-G_1(\vec v;\vec y))\Big)A_1^c(\vec z)A_1^e(\vec y)  \nn\\
&& + {1\over 2}f^{adc}f^{dbe}\int_{y,z}\Big( (G_2(v_1,u_2;\vec z)-G_2(\vec v;\vec z))(G_2(\vec z;\vec y)-G_2(\vec v;\vec y))\nn\\
&& \qquad+ (G_2(\vec u;\vec z)-G_2(u_1,v_2;\vec z))(G_2(\vec z;\vec y)-G_2(u_1,v_2;\vec y))\Big)A_2^c(\vec z)A_2^e(\vec y)  \nn\\
&& + {1\over 2}f^{adc}f^{dbe}\int_{y,z} (G_1(\vec u;\vec y)-G_1(v_1,u_2;\vec y))(G_2(v_1,u_2;\vec z)-G_2(\vec v;\vec z))A_1^c(\vec y) A_2^e(\vec z) \nn\\
&& + {1\over 2}f^{ade}f^{dbc}\int_{y,z} (G_2(\vec u;\vec z)-G_2(u_1,v_2;\vec z))(G_1(u_1,v_2;\vec y)-G_1(\vec v;\vec y))A_2^e(\vec z)A_1^c(\vec y) \Bigg\}  \qquad 
\nn
\\
&=& {C_A\over 8\sqrt{\pi }}\mu  \int_{z} \vec{A}^c(\vec z)\cdot\vec{A}^c(\vec z) + O\left(\mu^{-1}\right)
= {C_A\over 8\sqrt{\pi }}\mu  \int_{\slashed{p}} \vec{A}^c(\vec p)\cdot\vec{A}^c(-\vec p) + O\left(\mu^{-1}\right)
\,.
\eE

Combining Eqs.~(\ref{F2I}), (\ref{F2II}) and (\ref{F2III}) we obtain 
\be
\int_{\slashed{p}}\delta_\mu(\vec p)|{\vec p}| \left(\vec{A}^a(\vec{p})\cdot \frac{\delta F_{GL}^{(2,2)}[\vec A]}{\delta \vec{A}^a(\vec{p})}\right) 
=
-\left(N+{1\over8}\right) {C_A\over 2 \pi} \int_{\slashed{p}}   \frac{1}{|\vec{p}|}  (\vec{p}\times\vec{A}^a(-\vec{p})) (\vec{p}\times\vec{A}^a(\vec{p})) \,.
\ee
Note that the divergent term has disappeared on the right-hand-side so we can take the $\mu \rightarrow \infty$ limit. This equation can be solved using Eqs.~(20) and (21) of Ref.~\cite{Krug:2013yq}. We obtain
\bE{rCl}
F^{(2,2)}_{GL}[\vec A]  &=&   -\left(N+{1\over8}\right) {C_A\over 4 \pi} \int_{\slashed{p}}   \frac{1}{|\vec{p}|^2}  (\vec{p}\times\vec{A}^a(-\vec{p})) (\vec{p}\times\vec{A}^a(\vec{p})) \label{F22GL} \,.
\eE
This concludes the computation of the wave functional with ${\cal O}(e^2)$ precision. The complete result is summarized in Eqs.~(\ref{FGL0p}), (\ref{FGL1}), (\ref{F24GL}) and (\ref{F22GL}). Note that the result is different from the one obtained in Ref.~\cite{Krug:2013yq}. The reason is that the 
prefactor of Eq.~(\ref{F22GL}) has changed: $N \rightarrow N+1/8$. This highlights the importance of doing the regularization of the theory from the very beginning. The existence of very lengthy and complicated expressions in the intermediate steps impedes in practice the identification of the divergences. 
Therefore, one could easily miss some contributions (and yet get a finite result) if formally manipulating the integrals as if they were finite before regulating them.

\section{Determination of $\Psi_{GI}[J]$}
\label{sec:KKN}
In Refs.~\cite{Karabali:1995ps,Karabali:1996je,Karabali:1996iu, Karabali:1997wk,Karabali:1998yq,Karabali:2009rg} the \SE 
$ $
 was reformulated in terms of gauge invariant field variables named $J$. This has the great advantage that the Gauss law constraint is trivially satisfied. The original motivation of those works was to understand the strong coupling limit, but the approximation scheme worked out in Ref.~\cite{Karabali:2009rg} can be reformulated to provide with a systematic expansion of the weak coupling limit, and we did so in Ref.~\cite{Krug:2013yq}.\\
In order to arrive at the fields $J$, a series of field variable transformations has been used. First one defines the holomorphic and anti-holomorphic gauge fields
\be
\label{AbarA}
A:=\frac{1}{2}\left(A_1+iA_2\right) \quad \mathrm{and}\quad \bar A:=\frac{1}{2}\left(A_1-iA_2\right)
\,,
\ee
which makes it convenient to also change the space and momentum components to complex variables (note that $k$ and $z$ are defined with different signs):
\begin{IEEEeqnarray}{rClrCl}
z &=& x_1-ix_2, \qquad &\bar z &=& x_1+ix_2, \nn\\
k &=& \frac{1}{2}(k_1+ik_2),  &\bar k &=& \frac{1}{2}(k_1-ik_2), \quad {\vec k}\cdot {\vec x}=\bar k \bar z+kz, \\
\partial &=& \frac{1}{2}\left(\partial_1+i\partial_2\right), 
&\bar \partial &=& \frac{1}{2}\left(\partial_1-i\partial_2\right),  \quad \partial\bar\partial=\frac{1}{4}\vec\nabla^2\,,  \nn
\end{IEEEeqnarray}
Next one defines SL($N$,$\mathbb{C}$) matrices $M$ and $M^\dagger$ by
\be
A=-\frac{1}{e}(\partial M) M^{-1} \quad \mathrm{and}\quad \bar A=\frac{1}{e}M^{\dagger-1} (\bar\partial M^{\dagger}) \label{AofM}
\,,
\ee
the gauge invariant fields
\be
H=M^\dagger M \label{H}
\,,
\ee
and the gauge invariant currents
\be
J=\frac{2}{e}\p H H^{-1}\ =J^aT^a\,. \label{J}
\ee

A set of useful equalities for this section are relegated to App.~\ref{sec:useful}.

\subsection{Regulating the kinetic term}
One important consequence of this approach is that, since the vacuum wave functional is gauge invariant, it only depends on $J$. It is also 
possible to obtain an explicit and compact expression for the Hamiltonian in terms of $J$ fields. This was done
in Refs.~\cite{Karabali:1995ps,Karabali:1996je,Karabali:1996iu, Karabali:1997wk,Karabali:1998yq,Karabali:2009rg}, starting with a regularized Hamiltonian.
Interestingly enough, the regularization of the kinetic operator produced a finite extra term in the Hamiltonian. Yet, the expression found in those references will prove to be insufficient for our purposes. Therefore, as the regularization is an important point for us, we will rederive the Hamiltonian in terms of the $J$ fields. 
In several aspects the derivation will be identical to the one carried out in Refs.~\cite{Karabali:1995ps,Karabali:1996je,Karabali:1996iu, Karabali:1997wk,Karabali:1998yq,Karabali:2009rg},  but we will see that we need to consider some extra terms. 
Our starting point is the regularized kinetic operator  $\mathcal{T}_{reg}$ defined in Eq.~(\ref{Treg}). 
We then write the kinetic operator in terms of holomorphic and anti-holomorphic gauge fields\footnote{In Refs.~\cite{Karabali:1995ps,Karabali:1996je,Karabali:1996iu, Karabali:1997wk,Karabali:1998yq,Karabali:2009rg} the second term of Eq.~(\ref{TAbarA}) is not incorporated, but trivially considered to be equal to the first term. Yet, we find it illustrative to show their equality, as it is not evident from the actual computation after the change of variables.}:
\bE{l}
\label{TAbarA}
\T_{\mathrm{reg}}  = -\frac{1}{4}\int_{x,v}\delta_\mu(\vec x, \vec v) \Phi_{ab}(\vec x,\vec v)  \left(\frac{\delta}{\delta \bar{A}^a(\vec x)}\frac{\delta}{\delta A^b(\vec v)} + \frac{\delta}{\delta A^a(\vec x)} \frac{\delta}{\delta \bar{A}^b(\vec v)}\right) \,,
\eE
and transform it to $J$ variables. The functional derivatives of the first term can be rewritten in the following way
\bE{rCl}
&&
\frac{\delta}{\delta \bar{A}^a(\vec x)} \frac{\delta}{\delta A^b(\vec v)}  = \int_{y,z} \left[\frac{\delta J^d(\vec z)}{\delta \bar{A}^a(\vec x)} \frac{\delta}{\delta J^d(\vec z)} +\frac{\delta \bar{A}^d(\vec z)}{\delta \bar{A}^a(\vec x)} \frac{\delta}{\delta \bar{A}^d(\vec z)}  \right]    \left[\frac{\delta J^c(\vec y)}{\delta A^b(\vec v)} \frac{\delta}{\delta J^c(\vec y)} +\frac{\delta \bar{A}^c(\vec y)}{\delta A^b(\vec v)} \frac{\delta}{\delta \bar{A}^c(\vec y)}  \right] \quad
\\
&&\quad= \int_{y,z} \left[-2iM^\dagger_{dh}(\vec z)\left(D_z^{he}\left(\bar{D}^{-1}\right)^{ea}_{zx}\right) \frac{\delta}{\delta J^d(\vec z)} +\delta(\vec x-\vec z) \frac{\delta}{\delta \bar{A}^a(\vec z)}  \right]    \left[2iM^\dagger_{cb}(\vec y)\delta(\vec y-\vec v) \frac{\delta}{\delta J^c(\vec y)}\right]\,.\quad
\eE
using the equalities of App.~\ref{sec:useful}. Accordingly, we find
\bE{rCl}
&&
\hspace{-0.5cm} \Phi_{ab}(\vec x,\vec v) \frac{\delta}{\delta \bar{A}^a(\vec x)} \frac{\delta}{\delta A^b(\vec v)}  \nn\\
&=& 2i  \Phi_{ab}(\vec x,\vec v) \frac{\delta M^\dagger_{cb}(\vec v)}{\delta \bar{A}^a(\vec x)} \frac{\delta}{\delta J^c(\vec v)}     \nn\\
&& +  4\int_{z}  \Phi_{ab}(\vec x,\vec v) \left[\left(\p_z \bar{G}(z-x)\right)M^\dagger_{da}(\vec x) + \frac{ie}{2}\bar{G}(z-x) f^{edf} J^e(\vec z)M^{\dagger}_{fa}(\vec x) \right]  M^{\dagger}_{cb}(\vec v) \frac{\delta^2}{\delta J^d(\vec z) \delta J^c(\vec v)}   
\nn\\
&& +2i   \Phi_{ab}(\vec x,\vec v) M^\dagger_{cb}(\vec v)\frac{\delta^2}{\delta \bar{A}^a(\vec x) \delta J^c(\vec v)}
\,. \label{PhiAbarA}
\eE
The last term is proportional to the Gauss law operator $I^a = -i\bar{D}^{ab}\frac{\delta}{\delta \bar{A}^b}= -iM^{\dagger-1}_{ad}\bar{\p}\left(M^\dagger_{db}\frac{\delta}{\delta \bar{A}^b}\right)$ (see Ref.~\cite{Krug:2013yq}), which vanishes on physical wave functionals. For the other two terms we have to take care of the regularization. 
Using Eqs.~(\ref{Mf}) and (\ref{MdaggerOverAbar}) we can rewrite the first term of Eq.~(\ref{PhiAbarA}) in the following way 
\bE{rCl}
2i \Phi_{ab}(\vec x,\vec v) \frac{\delta M^\dagger_{cb}(\vec v)}{\delta \bar{A}^a(\vec x)}
&=& 2ie \Phi_{ab}(\vec x,\vec v) \frac{1}{\pi(v-x)} M^{\dagger-1}_{bd}(\vec v) f^{dch} M^{\dagger-1}_{ah}(\vec x) \label{splitdMdA} \\
&=:& 2ie V_{hd}(\vec x,\vec v) \frac{1}{\pi(v-x)} f^{dch}\,, \label{1stTermV}
\eE
where we defined
\be
V^{dc}(\vec x,\vec v) := M_{da}^\dagger(\vec x)\Phi^{ab}(\vec x,\vec v)M^{\dagger-1}_{bc}(\vec v)\,.
\ee
We now turn to the second term of the regulated kinetic operator, Eq.~(\ref{TAbarA}):
\bE{rCl}
\Phi_{ab}(\vec x,\vec v) \frac{\delta}{\delta A^a(\vec x)} \frac{\delta}{\delta \bar{A}^b(\vec v)}
&=& \int_{y,z} \Phi_{ab}(\vec x,\vec v)  \left[2iM^\dagger_{ca}(\vec y)\delta(\vec y-\vec x) \frac{\delta}{\delta J^c(\vec y)}\right] \nn\\
&&\quad \left[-2iM^\dagger_{dh}(\vec z)\left(D_z^{he}\left(\bar{D}^{-1}\right)^{eb}_{zv}\right) \frac{\delta}{\delta J^d(\vec z)} +\delta(\vec v-\vec z) \frac{\delta}{\delta \bar{A}^b(\vec z)}  \right]  \; 
\eE
\bE{l}
= 2i \Phi_{ab}(\vec x,\vec v) M^\dagger_{ca}(\vec x) \frac{\delta^2}{\delta J^c(\vec x) \delta \bar{A}^b(\vec v)}  \\
\quad + 4 \int_z \left[  \left(\p_z \bar{G}(z-v)\right) V^{cd}(\vec x,\vec v) + \frac{ie}{2}\bar{G}(z-v) f^{edf} J^e(\vec z) V^{cf}(\vec x,\vec v) \right] \frac{\delta^2}{\delta J^c(\vec x)\delta J^d(\vec z)}   \nn\\
\quad + 4\Phi_{ab}(\vec x,\vec v)  \int_z M^\dagger_{ca}(\vec x)  \frac{\delta}{\delta J^c(\vec x)}    \left[  \left(\p_z \bar{G}(z-v)\right)M^\dagger_{db}(\vec v) + \frac{ie}{2}\bar{G}(z-v) f^{edf} J^e(\vec z)M^{\dagger}_{fb}(\vec v) \right] \frac{\delta}{\delta J^d(\vec z)}\,.  \nn
\eE
Again, the first term is proportional to the Gauss law operator $I^a$. After renaming $v\leftrightarrow x$ (which can be done under the integral) and using $V^{ba}(v,x)=V^{ab}(x,v)$ the second term is identical to the second term of Eq.~(\ref{PhiAbarA}). The third term reads after application of the functional derivative
\bE{l}
2ie\Phi_{ab}(\vec x,\vec v) M^\dagger_{ca}(\vec x)  \bar{G}(x-v) f^{cdf} M^{\dagger}_{fb}(\vec v)  \frac{\delta}{\delta J^d(\vec x)} \,.    
\eE
As $\bar{G}(-x)=-\bar{G}(x)$, this expression is identical to Eq.~(\ref{splitdMdA}). 

Therefore, we find that both subterms of Eq.~(\ref{TAbarA}) are equal. Summing them up and multiplying by $\left(-{1\over4}\right)$ we obtain
the completely regularized kinetic term to all orders in perturbation theory
\bE{rCl}
\label{TregKKNallorders}
\T_\mathrm{reg}
&=& -2  \int_{x,v,z} \delta_{\mu}({\vec x},{\vec v}) \left((\p_z \delta^{df} + \frac{ie}{2} f^{dfa} J^a(\vec z)){\bar G}(z-x)\right)V_{fc}({\vec x},{\vec v})
{\delta \over \delta J^d (\vec z)} {\delta \over \delta J^c (\vec v)}  \nn\\
&&   -ie\int_{x,v} \delta_{\mu}({\vec x},{\vec v})V_{hd}({\vec x},{\vec v})f^{dch}{\bar G}(v-x){\delta \over \delta J^c (\vec v)}\,,
\eE
This is a pure function of $J$, since $V_{dc}(\vec x,\vec v)$ is a gauge invariant object, which makes it possible to rewrite it completely
 in terms of $J$.
The easiest way to proceed is to first consider an infinitesimal path with small ${\vec v}-{\vec x}$. 
By Taylor expansion one finds 
\bE{rCl}
V_{dc}({\vec x},{\vec v}) &=& \de_{dc} - (v-x){e\over2} J_{dc}(\vec x) +{\cal O}(|\vec x -\vec v|^2)\,,
\eE
where we used $J_{dc}=-if^{dce}J^e$. 
By composition of these infinitesimal paths we obtain
\be
V_{dc}({\vec x},{\vec v})=\left( \mathcal{P} e^{{e\over2}\int_{\mathcal{C}}\d z J(\vec z)}\right)_{dc} \,.
\ee
Note that the integration is over the holomorphic component only. $V_{dc}(\vec x,\vec v)$ depends on the path, though physical results should not. 
For illustration, we show the explicit expression for small $|\vec x -\vec v|$ for the specific combination of paths that we consider in this paper:
\bea
\nn
V_{dc}({\vec x},{\vec v})&=&\delta_{dc} +
 {e\over2}\left[
 (x-v) J_{dc}(\vec v)+\frac{(x-v)^2}{2}\partial J_{dc}(\vec v)+\frac{(x-v)(\bar x-\bar v)}{2}\bar \partial J_{dc}(\vec v)
 \right]
\\
&& 
+ {e^2\over4}\frac{(x-v)^2}{2} (J(\vec v)J(\vec v))_{dc}+
{\cal O}(|\vec x -\vec v|^3)
\,.
\label{Ve2}
\eea
The ${\cal O}(e|\vec x -\vec v|)$ and ${\cal O}(e^2|\vec x -\vec v|^2)$ terms are path independent but not the ${\cal O}(e|\vec x -\vec v|^2)$ terms.

\medskip

The kinetic operator $\T_\mathrm{reg}$ admits a Taylor expansion in powers of $e$. We are only interested to keep the terms that may contribute to the 
wave functional to $\mathcal{O} (e^2)$. We first consider the second term of Eq.~(\ref{TregKKNallorders}). Inserting Eq.~(\ref{Ve2}) in Eq.~(\ref{1stTermV}) we find 
\bE{rCl}
\label{premassterm}
2i \Phi_{ab}(\vec x,\vec v) \frac{\delta M^\dagger_{cb}(\vec v)}{\delta \bar{A}^a(\vec x)}
&=& -\frac{e^2C_A}{\pi} J^c(\vec x) +{\cal O}(e^2|\vec x -\vec v|,e^3|\vec x -\vec v|) \,. \label{1stTermJ}
\eE
Note that regularization is crucial for obtaining a finite contribution, as the leading term from the Wilson line
(proportional to $\delta_{ab}$) vanishes.
Therefore, 
the integration of the regularized delta function times Eq.~(\ref{premassterm}) over $v$ gives
\bE{rCl}
\label{massterm}
-\frac{2}{4}\int_{x,v} \delta_\mu(\vec x, \vec v) 2i \Phi_{ab}(\vec x,\vec v) \frac{\delta M^\dagger_{cb}(\vec v)}{\delta \bar{A}^a(\vec x)} \frac{\delta}{\delta J^c(\vec v)}
&=&  \frac{e^2C_A}{2 \pi}\int_x J^c(\vec x) \frac{\delta}{\delta J^c(\vec x)}+{\cal O}(e^2/\mu,e^3/\mu) \,.
\eE
This contribution to the kinetic operator has been generated by the regularization of the theory, i.e.~it is an effect produced by the high-energy modes. 
It was first obtained in Ref.~\cite{Karabali:1996je}, and it has a nice interpretation in terms of an anomaly-like computation. This term 
has played a major role in the strong coupling analysis carried out in Refs.~\cite{Karabali:1995ps,Karabali:1996je,Karabali:1996iu, Karabali:1997wk,Karabali:1998yq,Karabali:2009rg}, 
where it is argued to be responsible for generating the mass gap. 
Yet, we would like to remark, as it is clear from the 
analysis above, that this contribution is obtained from a purely perturbative computation (as anomaly-like effects are anyway), arising from a Taylor 
expansion in powers of $e$. The corrections to this expression are $1/\mu$ suppressed, irrespectively of the power of $e$ (but starting at ${\cal O}(e^2)$). In general we may worry that such $1/\mu$ suppression may be compensated by divergences when applied to the wave functional. This is not the case for this term, as there is a complete factorization between the momentum of the internal loop and the momentum of the fields that will act on the wave functional. Therefore, we will not consider these vanishing contributions explicitly any further in this paper (even though they are formally ${\cal O}(e^2)$). 

\medskip

We now move to the first term of Eq.~(\ref{TregKKNallorders}). The expansion of $V$ around $v=x$ yields
 \bE{rCl}
 \label{OmReg} 
&=& -2\int_{z}  \Bigg[\left(\p_z \bar{G}(z-x)\right)\de^{dc} + \frac{ie}{2}\bar{G}(z-x) f^{dce} J^e(\vec z) \\
&&\qquad +(v-x)\frac{ie}{2} \left(\p_z \bar{G}(z-x)\right)f^{dce}J^e(\vec x) \nn\\
&&\qquad - (v-x) \frac{e^2}{8}f^{dea}f^{ecb} \Big( (v-x)\left(\p_z \bar{G}(z-x)\right) J^a(\vec x) +2\bar{G}(z-x) J^a(\vec z) \Big) J^b(\vec x)  \Bigg] \frac{\delta^2}{\delta J^d(\vec z) \delta J^c(\vec v)}\,.\qquad 
\nn
\eE
The last two lines are of ${\cal O}(e|\vec x -\vec v|)$ and ${\cal O}(e^2|\vec x -\vec v|^2)$ respectively, but when applied to a functional they can give finite contributions.
We have not included ${\cal O}(e|\vec x -\vec v|^2)$ terms in this expansion. In principle they may 
contribute to the wave functional at ${\cal O}(e^2)$. Nevertheless, as we will see in the following, only the ${\cal O}(e|\vec x -\vec v|)$ terms give 
finite contributions at ${\cal O}(e^2)$. Therefore, the ${\cal O}(e|\vec x -\vec v|^2)$ terms would give, at most, ${\cal O}(e^2/\mu)$ corrections to the wave functional. 
In order to maintain the expressions in a manageable way, we will neglect them in the following.

\medskip

After this discussion we can approximate the kinetic operator by an expression suitable to obtain the wave functional with ${\cal O}(e^2)$ accuracy:
\bE{rCl}
\label{TregKKN}
\T_\mathrm{reg}
&=& \frac{e^2C_A}{2\pi}  \int_x J^a (\vec x) {\delta \over \delta J^a (\vec x)}  + {2\over \pi} \int _{x,y} 
 {1\over (y-x)^2} {\delta \over \delta J^a (\vec x)} {\delta \over \delta J^a (\vec y)} + i e \int_{x,y} f^{abc} {J^c(\vec x) \over \pi (y-x)} {\delta \over \delta J^a (\vec x)} {\delta \over \delta J^b (\vec y)}  \nn\\
&&   +\int_{x,v,y} \delta_\mu(\vec x, \vec v)  \Bigg[(x-v)ie \left(\p_y \bar{G}(y-v)\right)f^{abe} J^e(\vec v) \nn\\
&&\qquad + (x-v) \frac{e^2}{4}f^{ace}f^{bed} J^c(\vec v) \Big( (x-v)\left(\p_y \bar{G}(y-v)\right) J^d(\vec v) +2\bar{G}(y-v) J^d(\vec y) \Big)   \Bigg] \frac{\delta^2}{\delta J^a(\vec x)\delta J^b(\vec y)}  \nn\\
 && -\int_{y,z} \bar{G}(y-z) M^\dagger_{ca}(\vec y) {\delta \over \delta J^c (\vec z)} I^a(\vec y) \\
 &&+{\cal O}(e^3,1/\mu) \nn\\
&=:& \int_x\omega(\vec x)^a {\delta \over \delta J^a (\vec x)} + \int_{x,v,y}\tilde{\Omega}_{ab}^\mathrm{reg}(\vec x,\vec v,\vec y)\frac{\delta^2}{\delta J^a(\vec x) \delta J^b(\vec y)} +{\cal O}(e^3) \label{omega} \\
&=:& \int_x\omega(\vec x)^a {\delta \over \delta J^a (\vec x)} \nn\\
&&+  \int_{x,v,y}\left(\Omega_{ab}^{(0)}(\vec x,\vec y) +e\Omega_{ab}^{(1)}(\vec x,\vec y) +e\tilde{\Omega}_{ab}^{(1)}(\vec x,\vec v,\vec y) +e^2\tilde{\Omega}_{ab}^{(2)}(\vec x,\vec v,\vec y) \right) \frac{\delta^2}{\delta J^a(\vec x) \delta J^b(\vec y)} 
\nn
\\
&&
+{\cal O}(e^3,1/\mu)  \,, \qquad \label{omega2}
\eE
where we dropped the term proportional to the Gauss law operator in the last two equalities, and we defined $\Omega_{ab}^{(0)}(\vec x,\vec y)$ and $\Omega_{ab}^{(1)}(\vec x,\vec y)$ as the coefficients of the second and the third term of the first line of Eq.~(\ref{TregKKN}), respectively, while $\tilde{\Omega}_{ab}^{(1)}(\vec x,\vec v,\vec y)$ and $\tilde{\Omega}_{ab}^{(2)}(\vec x,\vec v,\vec y)$ are the coefficients of the second and third line, respectively.

Eq.~(\ref{TregKKN}) agrees with the expression used in Ref. \cite{Karabali:2009rg} in  the limit $\mu \rightarrow \infty$. In this case they agree to any order in perturbation theory.
Nevertheless, as we will see, this is not enough for our purposes, and we will also have to keep some subleading terms in $1/\mu$.

\subsection{Solving the \SE}

Once we have obtained the regulated kinetic operator we can compute the $\Psi_{GI}$. 
After changing to the $J$ variables Eq. (\ref{SE}) reads in our case
\be
\V-\int_{x}
\omega^a(\vec x)\frac{\delta F_{GI}}{\delta J^a(\vec x)}-
\int_{x,v,y}\tilde{\Omega}^\mathrm{reg}_{ab}(\vec x,\vec v,\vec y)\frac{\delta^2 F_{GI}}{\delta J^a(\vec x) \delta J^b(\vec y)}  + \int_{x,v,y}\tilde{\Omega}^\mathrm{reg}_{ab}(\vec x,\vec v,\vec y)\frac{\delta F_{GI}}{\delta J^a(\vec x)} \frac{\delta F_{GI}}{\delta J^b(\vec y)} = 0 \,, \label{SEJ}
\ee
where
\be
\V=\frac{1}{2}\int_x{\bar \partial}J^a(\vec x){\bar \partial}J^a(\vec x)
\,,
\ee
and $\omega^a(\vec x)$ and $\tilde{\Omega}^\mathrm{reg}_{ab}(\vec x,\vec v,\vec y)$ are defined in Eq.~(\ref{omega}).
As before we expand the exponent of the vacuum wave functional in powers of the coupling constant
\be
F_{GI}=F_{GI}^{(0)}+eF_{GI}^{(1)}+e^2F_{GI}^{(2)}+{\cal O}(e^3)\,,
\ee
and separate the \SE $ $ order by order in the coupling constant.

At ${\cal O}(e^0)$ we have
\bE{l}
\int_{x,y}\Omega^{(0)}_{ab}(\vec x,\vec y) \left(\frac{\delta^2 F_{GI}^{(0)}}{\delta J^a(\vec x) \delta J^b(\vec y)} - \frac{\delta F_{GI}^{(0)}}{\delta J^a(\vec x)} \frac{\delta F_{GI}^{(0)}}{\delta J^b(\vec y)} \right)= {1\over 2} \int_z  \bar{\p} J^a(\vec z) \bar{\p} J^a(\vec z)  \,. 
\eE 
This, as before, is the unregularized lowest order \SE. Its solution is the leading order computed in Ref.~\cite{Krug:2013yq}. It corresponds to the weak coupling limit of the leading order of Ref.~\cite{Karabali:2009rg}:
\bE{l}
\label{FGI0}
F^{(0)}_{GI} ={1\over2} \int_\slashed{k} \frac{\bar{k}^2}{E_k} J^a(\vec{k}) J^a(-\vec{k})   = {1\over2} \int_\slashed{k}\frac{1}{E_k} (\vec{k}\times\vec{A}^a(\vec{k})) (\vec{k}\times\vec{A}^a(-\vec{k}))+ {\mathcal O}(e) 
 = F_{GL}^{(0)}[{\vec A}] + {\mathcal O}(e) \,, \qquad 
\eE
 where $E_k \equiv |\vec k|$.

At ${\cal O}(e^1)$ we have
\bE{l}
-\int_{x,y}\Omega^{(0)}_{ab}(\vec x,\vec y) \left(\frac{\delta^2 F_{GI}^{(1)}}{\delta J^a(\vec x) \delta J^b(\vec y)} - 2 \frac{\delta F_{GI}^{(0)}}{\delta J^a(\vec x)} \frac{\delta F_{GI}^{(1)}}{\delta J^b(\vec y)} \right)\nn\\
\quad -\int_{x,y} \Omega^{(1)}_{ab}(\vec x,\vec y)\left(\frac{\delta^2 F_{GI}^{(0)}}{\delta J^a(\vec x) \delta J^b(\vec y)} - \frac{\delta F_{GI}^{(0)}}{\delta J^a(\vec x)} \frac{\delta F_{GI}^{(0)}}{\delta J^b(\vec y)} \right) \nn\\
\quad -\int_{x,v,y}\tilde{\Omega}^{(1)}_{ab}(\vec x,\vec v,\vec y) \left(\frac{\delta^2 F_{GI}^{(0)}}{\delta J^a(\vec x) \delta J^b(\vec y)} - \frac{\delta F_{GI}^{(0)}}{\delta J^a(\vec x)} \frac{\delta F_{GI}^{(0)}}{\delta J^b(\vec y)} \right) = 0 \,. 
\eE 
The first term of the last line vanishes under contraction of the color indices. The second term is of $O(\mu^{-2})$ (see App.~\ref{subsec:append1}). So, like for the leading order, this equation reduces to the unregularized version. Thus, its solution is the one quoted in Ref.~\cite{Krug:2013yq}, which also corresponds to the ${\cal O}(e)$ weak coupling limit of the solution given in 
Ref.~\cite{Karabali:2009rg}:
\be
F^{(1)}_{GI}=-\frac{1}{4}\int_{\slashed{k_1},\slashed{k_2},\slashed{k_3}} \frac{f^{a_1 a_2 a_3}}{24} \sd (\vec k_1+\vec k_2+\vec k_3)\  g^{(3)}(\vec k_1,\vec k_2,\vec k_3) J^{a_1}(\vec k_1)J^{a_2}(\vec k_2)J^{a_3}(\vec k_3)\,, \label{FGI1}
\ee
where
\be
g^{(3)}(\vec k_1,\vec k_2,\vec k_3) = \frac{16}{E_{k_1}\! + E_{k_2}\! + E_{k_3}}\left \{ \frac{\bar k_1 \bar k_2 (\bar k_1 - \bar k_2)}{E_{k_1} E_{k_2}} + {cycl.\ perm.} \right \}
\,. 
\ee

At ${\cal O}(e^2)$ we determine $F_{GI}^{(2)}$. As in the previous section, $F^{(2)}_{GI}$ can have contributions with four, two and zero $J$'s:
$F^{(2)}_{GI}=F^{(2,4)}_{GI}+F^{(2,2)}_{GI}+F^{(2,0)}_{GI}$. Again, there is no need to compute $F^{(2,0)}_{GI}$, as it only changes the normalization of the state, which we do not fix, or alternatively can be absorbed in a redefinition of the ground-state energy. $F^{(2,4)}_{GI}$ is determined by the following equation (where $\Omega^{(1)}_{ab}(\vec x,\vec y)$ and $\tilde{\Omega}^{(1)}_{ab}(\vec x,\vec v,\vec y)$ should be understood in a symmetrized way):
\bE{l}
 \int_{x,y}\Omega^{(0)}_{ab}(\vec x,\vec y)\left(\frac{\delta F_{GI}^{(1)}}{\delta J^a(\vec x)} \frac{\delta F_{GI}^{(1)}}{\delta J^b(\vec y)}  + 2\frac{\delta F_{GI}^{(0)}}{\delta J^a(\vec x)} \frac{\delta F_{GI}^{(2,4)}}{\delta J^b(\vec y)} \right)  +2\int_{x,y}\Omega^{(1)}_{ab}(\vec x,\vec y)\frac{\delta F_{GI}^{(0)}}{\delta J^a(\vec x)} \frac{\delta F_{GI}^{(1)}}{\delta J^b(\vec y)}  \nn\\
\quad  +2\int_{x,v,y}\tilde{\Omega}^{(1)}_{ab}(\vec x,\vec v,\vec y)\frac{\delta F_{GI}^{(0)}}{\delta J^a(\vec x)} \frac{\delta F_{GI}^{(1)}}{\delta J^b(\vec y)}  +\int_{x,v,y}\tilde{\Omega}^{(2)}_{ab}(\vec x,\vec v,\vec y)\frac{\delta F_{GI}^{(0)}}{\delta J^a(\vec x)} \frac{\delta F_{GI}^{(0)}}{\delta J^b(\vec y)}  = 0 \,, \qquad
\eE 
When $\mu\to\infty$ the last line vanishes (see App.~\ref{subsec:append2}), and the equation reduces to the unregularized equation with the solution
\be
\label{FGI24}
F^{(2,4)}_{GI}=-\frac{1}{8}\int_{\slashed{k_1},\slashed{k_2},\slashed{q_1},\slashed{q_2}} \frac{f^{a_1 a_2 c} f^{b_1 b_2 c}}{64} \sd (\vec k_1+\vec k_2+\vec q_1+\vec q_2)  g^{(4)}(\vec k_1, \vec k_2; \vec q_1, \vec q_2) J^{a_1}(\vec k_1)J^{a_2}(\vec k_2)J^{b_1}(\vec q_1)J^{b_2}(\vec q_2)\,,
\ee
where
\begin{equation}
\begin{array}{cl}
g^{(4)}(\vec k_1, \vec k_2; \vec q_1, \vec q_2)& =\ \vspace{.2in} \displaystyle \frac{1}{E_{k_1}\! + E_{k_2}\! + E_{q_1}\! + E_{q_2}} \\
\vspace{.2in}
&\!\!\!\!\displaystyle \left \{ g^{(3)}(\vec k_1, \vec k_2, -\vec k_1-\vec k_2)\ \frac{k_1 + k_2}{\bar k_1 +\bar k_2}\ g^{(3)}(\vec q_1, \vec q_2, -\vec q_1-\vec q_2) \right . \\
\vspace{.2in}
&\displaystyle -  \left [ \frac{(2\bar k_1 + \bar k_2)\,\bar k_1}{E_{k_1}} - \frac{(2\bar k_2 + \bar k_1)\,\bar k_2}{ E_{k_2}}\right ]\frac{4}{\bar k_1+\bar k_2}\  g^{(3)}(\vec q_1, \vec q_2, -\vec q_1-\vec q_2) \\
&\displaystyle -  \left .   g^{(3)}(\vec k_1, \vec k_2, -\vec k_1-\vec k_2)\ \frac{4}{\bar q_1+\bar q_2}\left [ \frac{(2\bar q_1 + \bar q_2)\,\bar q_1}{ E_{q_1}} - \frac{(2\bar q_2 + \bar q_1)\,\bar q_2}{ E_{q_2}}\right ] \right\} \,.
\\
\end{array}
\end{equation}
Again, this term corresponds to the weak coupling limit of the the analogous expression in Ref.~\cite{Karabali:2009rg}, and to the expression already found in Ref.~\cite{Krug:2013yq}.

So far the regularization of the kinetic term has not produced any modification of the results obtained in Ref.~\cite{Krug:2013yq}. 
The reason is the same as in the previous section: so far all computations we did were tree-level-like. ``Loop" effects (sensitive to the hard modes) are hidden in $F^{(2,2)}_{GI}$, where we have a kind of contraction of two fields. We compute this term in the 
next subsection.
 
 \subsubsection{$F^{(2,2)}_{GI}$}
 $F^{(2,2)}_{GI}$ is determined by the following equation
\bE{l}
-{C_A\over2\pi}\int_x J^a(\vec x)\frac{\delta F_{GI}^{(0)}}{\delta J^a(\vec x)} - \int_{x,y}\Omega^{(0)}_{ab}(\vec x,\vec y)\left(\frac{\delta^2 F_{GI}^{(2,4)}}{\delta J^a(\vec x) \delta J^b(\vec y)}  - 2\frac{\delta F_{GI}^{(0)}}{\delta J^a(\vec x)} \frac{\delta F_{GI}^{(2,2)}}{\delta J^b(\vec y)} \right) \nn\\
\quad  -\int_{x,y}\Omega^{(1)}_{ab}(\vec x,\vec y) \frac{\delta^2 F_{GI}^{(1)}}{\delta J^a(\vec x) \delta J^b(\vec y)}  \nn\\
\quad  -\int_{x,v,y}\tilde{\Omega}^{(1)}_{ab}(\vec x,\vec v,\vec y)\frac{\delta^2 F_{GI}^{(1)}}{\delta J^a(\vec x) \delta J^b(\vec y)}  -\int_{x,v,y}\tilde{\Omega}^{(2)}_{ab}(\vec x,\vec v,\vec y)\frac{\delta^2 F_{GI}^{(0)}}{\delta J^a(\vec x) \delta J^b(\vec y)}  = 0 \,.
\label{SEF22}
\eE 
The last term vanishes in the $\mu\to\infty$ limit (see App.~\ref{subsec:append3}), the next-to-last term, however, does not. 
With Eqs.~(\ref{omega2}) and (\ref{FGI1}) we find
\bE{rl}
 & \int_{x,v,y}\tilde{\Omega}^{(1)}_{ab}(\vec x,\vec v,\vec y)\frac{\delta^2 F_{GI}^{(1)}}{\delta J^a(\vec x) \delta J^b(\vec y)}   \nn\\
= & 3{C_A\over48\mu^2} \int_{\slashed{k},\slashed{p}} \frac{p(\bar{k}+\bar{p})}{\bar{p}} e^{-\frac{(\vec{k}+\vec{p})^2}{4\mu^2}} g^{(3)}(\vec p,\vec k,-\vec k-\vec p)  J^a(-\vec k)J^a(\vec k)\,.
\eE
In order to compute the loop integral over the internal ${\vec p}$ momentum, we again factorize the modes according to the two scales of the problem: $|{\vec p}| \sim \mu$ and $|{\vec p}| \sim |{\vec k}|$.
The integral is dominated by $|{\vec p}| \sim \mu$, while the $|{\vec p}| \sim |{\vec k}|$ region gives subleading contributions.
Overall we obtain (here $\alpha$ is the angular component of $\vec{k}$, such that $\bar{k}={1\over2}|\vec{k}|e^{-i\alpha}$):
\bea
\label{new}
&&
\int_{x,v,y}\tilde{\Omega}^{(1)}_{ab}(\vec x,\vec v,\vec y)\frac{\delta^2 F_{GI}^{(1)}}{\delta J^a(\vec x) \delta J^b(\vec y)}  
\\
\nn
&=&
{C_A\over16\mu^2(2\pi)^2} \int_{\slashed{k}} \Big(\frac{9}{4} e^{-2 i \alpha } |\vec{k}|^2 \pi ^{3/2} \mu -7 e^{-2 i \alpha } |\vec{k}| \pi  \mu ^2\Big) 
J^a(-\vec k)J^a(\vec k) +{\cal O}(1/\mu^{2})
 \\
\nn
 &=&
 -{7\over8}{C_A\over2\pi} \int_{\slashed{k}} \frac{\bar{k}^2}{|\vec{k}|}  J^a(-\vec k)J^a(\vec k) +{\cal O}(1/\mu)\,.
\eea

We now have all the ingredients to determine $f^{(2,2)}_{a_1a_2}(k)$ from Eq.~(\ref{SEF22}), which now reads:
\bE{l}
{C_A\over2\pi}\int_x J^a(\vec x)\frac{\delta F_{GI}^{(0)}}{\delta J^a(\vec x)} + \int_{x,y}\Omega^{(0)}_{ab}(\vec x,\vec y)\left(\frac{\delta^2 F_{GI}^{(2,4)}}{\delta J^a(\vec x) \delta J^b(\vec y)}  - 2\frac{\delta F_{GI}^{(0)}}{\delta J^a(\vec x)} \frac{\delta F_{GI}^{(2,2)}}{\delta J^b(\vec y)} \right) \nn\\
\quad  +\int_{x,y}\Omega^{(1)}_{ab}(\vec x,\vec y) \frac{\delta^2 F_{GI}^{(1)}}{\delta J^a(\vec x) \delta J^b(\vec y)}  -{7\over8}{C_A\over2\pi} \int_{\slashed{k}} \frac{\bar{k}^2}{|\vec{k}|}  J^a(-\vec k)J^a(\vec k)   = 0  \qquad
\eE 
\bE{rCl}
\Longleftrightarrow 2\int_{\slashed{k}} |\vec{k}|  f^{(2,2)}_{a_1a_2}(\vec k) J^{a_1}(-\vec k)J^{a_2}(\vec k) &=& -\frac{C_A}{32} \int_{\slashed{k},\slashed{p}}\frac{p}{\bar{p}}  g^{(4)}(\vec k,\vec p,-\vec k,-\vec p)  J^a(-\vec k)J^a(\vec k)  \nn\\
&&  -\frac{C_A}{16} \int_{\slashed{k},\slashed{p}} \frac{1}{\bar{p}} g^{(3)}(\vec k,\vec p,-\vec p-\vec k) J^a(-\vec k)J^a(\vec k)  \nn\\
&&  -\left(1-{7\over8}\right){C_A\over2\pi} \int_{\slashed{k}} \frac{\bar{k}^2}{|\vec{k}|} J^a(-\vec k)J^a(\vec k)\,,   \qquad
\eE 
and it is solved by
\bE{rCl}
 f^{(2,2)}_{a_1a_2}(\vec k) &=&- {C_A\over4\pi} \Bigg( N  +{1\over8}\Bigg) \frac{\bar{k}^2}{|\vec{k}|^2} \de_{a_1a_2} \,,
\eE 
where $N= 0.025999\,(8\pi) $ was defined in Eq.~(\ref{N}). Therefore, $e^2F_{GI}^{(2,2)}$ reads
\bE{rCl}
\label{FGI22}
 e^2F_{GI}^{(2,2)} &=& -\Bigg( N +{1\over8}\Bigg)  {e^2C_A\over4\pi} \int_\slashed{k} \frac{\bar{k}^2}{|\vec{k}|^2}  J^a(-\vec k)J^a(\vec k) \\
 &=& -\Bigg( N +{1\over8}\Bigg)  {e^2C_A\over4\pi} \int_\slashed{k} \frac{1}{|\vec{k}|^2}  (\vec{k}\times\vec{A}^a(-\vec{k})) (\vec{k}\times\vec{A}^a(\vec{k})) +{\cal O}(e^3) \\
  &=& e^2F^{(2,2)}_{GL} +{\cal O}(e^3) \nn\,.
\eE 
This concludes the computation of the wave functional with ${\cal O}(e^2)$ precision in terms of $J$ fields. The complete result is summarized in Eqs.~(\ref{FGI0}), (\ref{FGI1}), (\ref{FGI24}) and (\ref{FGI22}).
This result differs from the expression obtained in Ref.~\cite{Krug:2013yq}, and from the weak coupling limit of the expression obtained in Ref.~\cite{Karabali:2009rg}. 
The reason is that the 
prefactor of Eq.~(\ref{FGI22}) has changed: $N +1 \rightarrow N+1/8$. This is important, as now the new prefactors of Eqs.~(\ref{F22GL})  and
(\ref{FGI22}) agree with each other. This was the missing ingredient to claim complete agreement between both computations, which now we do:
The vacuum wave functional computed with methods (A) and (B) agree with each other with ${\cal O}(e^2)$ precision (when written with the same variables, either $J$ or ${\vec A}$). In other words
\be
F^{(0)}_{GI} + e F^{(1)}_{GI} + e^2 (F^{(2,2)}_{GI}+F^{(2,4)}_{GI}) = F^{(0)}_{GL} + e F^{(1)}_{GL} + e^2 (F^{(2,2)}_{GL}+F^{(2,4)}_{GL})+{\cal O}(e^3)\,.
\ee

Finally, let us note that  the ``mass term" Eq.~(\ref{massterm}), which is taken to be responsible for generating the mass gap in strong coupling analysis, is not a special term from the point of view of weak coupling, as there are more terms in the Hamiltonian Eq.~(\ref{TregKKN}) that produce identical terms to the wave functional (see, for instance, Eq.~(\ref{new})).

\section{Conclusions}

We have obtained the complete expression for the Yang-Mills vacuum wave functional in three dimensions at weak coupling with ${\cal O}(e^2)$ precision. We have used two different methods to solve the Schr\"odinger functional equation: 
(A) One of them generalizes to ${\cal O}(e^2)$ the method followed by Hatfield at ${\cal O}(e)$~\cite{Hatfield:1984dv}. We have named the 
result $\Psi_{GL}[{\vec A}]$. 
(B) The other uses the weak coupling version of the gauge invariant formulation of the Schr\"odinger equation and the ground-state wave functional followed by Karabali, Nair, and Yelnikov \cite{Karabali:2009rg}. We have named the result $\Psi_{GI}[J]$. 
Such computations had been addressed previously in Ref.~\cite{Krug:2013yq} 
obtaining conflicting results between both methods. 
Nevertheless, possible new effects associated to the regularization of the Hamiltonian were not studied. 
Such study has been carried out in full detail in this paper. 
This has led in both cases to new (but different) contributions emanating from the regularization of the theory. 
The final results for both methods now agree with each other. This is a 
very strong check of the computations and of the regularization procedure used in this paper. We can now claim that we have obtained the complete 
expression of the Yang-Mills vacuum wave functional in three dimensions with ${\cal O}(e^2)$ precision for the first time. In terms of the ${\vec A}$ fields the vacuum wave functional can be found in Eqs.~(\ref{FGL0p}), (\ref{FGL1}), (\ref{F24GL}) and (\ref{F22GL}),  and in terms of the gauge invariant $J$ variable in Eqs.~(\ref{FGI0}), (\ref{FGI1}), (\ref{FGI24}) and (\ref{FGI22}). Both results are equal to ${\cal O}(e^2)$. To our knowledge this is the first time that a full fledge (including regularization) computation of the wave functional of a gauge theory has been undertaken.

The fact that the result obtained in this paper differs from the one obtained in Ref.~\cite{Krug:2013yq} with method (A) should not be so surprising, 
as the regularization of the kinetic operator was not considered there. More surprising is the fact that a new term, Eq.~(\ref{new}), has been found 
using method (B), the regularization of which had been studied in detail in the past, albeit in the strong coupling limit  
 (see, for instance, the discussion in Refs.~\cite{Karabali:1997wk,Agarwal:2007ns}, in particular in the Appendix of the last reference). In those references an intermediate cutoff $\mu'  \ll \mu $ was introduced in the wave functional damping its high energy modes (compared with $\mu$). Such procedure kills the extra contribution we have found with method (B) in this paper. Nevertheless, it also eliminates the mass term obtained with method (A), producing two different results in our computation. Instead, we advocate that the whole computation should be 
 done with a single cutoff $\mu$ that regulates the kinetic term and then the ground state wave functional (and all excitations) at the same time. It is only after solving the Schr\"odinger equation that we can take the cutoff $\mu$ to infinity compared with any finite momentum of the system. As one goes to higher orders in perturbation theory loops appear, whose integrals run up to infinity. In other words, the momentum of the fields of the wave functional can be large (in loops), producing new contributions, as we have seen in this paper (see Eq.~(\ref{new})). In any case it is clear that regularization of the wave functional in the Schr\"odinger formalism is still in its infancy, and more work is needed to put the formalism on more solid ground. In this respect we would like to mention that an additional check of our wave functional could be the computation (and comparison with known results) of the static potential at ${\cal O}(e^2)$ from the expectation value of the Wilson loop. We also want to explore in 
the future the consequences of our results for the approximated resummation scheme analysis carried out in Ref.~\cite{Karabali:2009rg}. 

Finally, we cannot avoid making some considerations of the possible significance of the mass-like term (\ref{FGI22}). Its mass prefactor is gauge independent. Following Refs.~\cite{Karabali:1995ps,Karabali:1996je,Karabali:1996iu, Karabali:1997wk,Karabali:1998yq} 
one may argue about its relation with the magnetic screening mass. If we do so we obtain
\be
m=\left(\frac{1}{8} + (8\pi)0.025999\right)\frac{C_A e^2}{2\pi}=0.778426\frac{C_A e^2}{2\pi}=0.247781\frac{C_A }{2}e^2
\,.
\ee
This value is in the same ballpark as the values obtained from some resummation schemes of perturbation theory at one loop \cite{Alexanian:1995rp,Jackiw:1995nf,Buchmuller:1996pp,Cornwall:1997dc}\footnote{At two loops the result depends on the renormalization scale, see Table I of Ref.~\cite{Bieletzki:2012rd}, but the agreement is still reasonable.}. In particular, it is remarkable close to the value quoted in Ref.~\cite{Cornwall:1997dc}. It is also not far from the mass value proposed in Ref.~\cite{Karabali:1995ps}: $m=\frac{C_A e^2}{2\pi}$, which was obtained from a strong coupling computation at leading order.

\bigskip

\acknowledgments{
We acknowledge discussions with D. Karabali and V.P. Nair. This work was partially supported by the Spanish 
grants FPA2010-16963 and FPA2011-25948, and by the Catalan grant SGR2009-00894.
}

\vfill
\newpage

\appendix

\section{Functional derivative of the string}
\label{sec:herm}
We use Eq.~(\ref{M_iOverA_j}) to compute
\be
\half\sum_i\int_{x,v}\de_\mu(\vec x,\vec v) \left[\frac{\de}{\de A_i^a(\vec x)}\Phi_{ab}(\vec x,\vec v)\right]\frac{\de}{\de A_i^b(\vec v)} 
\ee
This is actually an ill-defined quantity, so we have to regularize it. We do this by moving the derivative an infinitesimal step $\vec X$ away from the point $\vec x$ and introduce a new regulated delta function and a second string. We then take the limit $\nu \rightarrow \infty$ for finite $\mu$. 
\bE{ll}
\half&\lim_{\nu\to\infty}\sum_i\int_{x,v,X}\de_\mu(\vec x,\vec v)\de_\nu(\vec X)\Phi_{ar}(\vec x,\vec x+\vec X)\left[\frac{\de}{\de A_i^r(\vec x+\vec X)}\Phi_{ab}(\vec x,\vec v)\right]\frac{\de}{\de A_i^b(\vec v)} \nn\\
&={1\over4}\lim_{\nu\to\infty} \sum_i\int_{x,v}\de_\mu(\vec x,\vec v) \de_\nu(\vec X)\Phi_{ar}(\vec x,\vec x+\vec X)\frac{\de}{\de A_i^r(\vec x+\vec X)} \Bigg[M_1(\vec x)M_1^{-1}(v_1,x_2)M_2(v_1,x_2)M_2^{-1}(\vec v) \nn\\
&\qquad \qquad +M_2(\vec x)M_2^{-1}(x_1,v_2)M_1(x_1,v_2)M_1^{-1}(\vec v)\Bigg]^{ab}\frac{\de}{\de A_i^b(\vec v)} \qquad\\
&={e\over4}\lim_{\nu\to\infty} \sum_i\int_{x,v} \de_\mu(\vec x,\vec v) \de_\nu(\vec X)\Phi_{ar}(\vec x,\vec x+\vec X) \nn\\
&\qquad\times\Bigg[\de_{i1}M_1^{ag}(\vec x)G_1(-\vec{X}) f^{gch} M_1^{rh}(\vec x+\vec X) [M_1^{-1}(v_1,x_2)M_2(v_1,x_2)M_2^{-1}(\vec v) ]^{cb} \nn\\
&\qquad \qquad -[M_1(\vec x)M_1^{-1}(v_1,x_2)]^{ac} \de_{i1}M_1^{cg}(v_1,x_2)G_1((v_1,x_2)-\vec{x}-\vec X) f^{gdh} M_1^{rh}(\vec x +\vec X) \nn\\
&\qquad \qquad\qquad \times [M_1^{-1}(v_1,x_2)M_2(v_1,x_2)M_2^{-1}(\vec v)]^{db} \nn\\
&\qquad \qquad +[M_1(\vec x)M_1^{-1}(v_1,x_2)]^{ac} \de_{i2}M_2^{cg}(v_1,x_2)G_2((v_1,x_2)-\vec{x}-\vec X) f^{gdh} M_2^{rh}(\vec x+\vec X) [M_2^{-1}(\vec v)]^{db} \nn\\
&\qquad \qquad -[M_1(\vec x)M_1^{-1}(v_1,x_2)M_2(v_1,x_2)M_2^{-1}(\vec v)]^{ac} \de_{i2}M_2^{cg}(\vec v)G_2(\vec v-\vec{x}-\vec X) f^{gdh} M_2^{rh}(\vec x+\vec X) [M_2^{-1}(\vec v)]^{db} \nn\\ 
&\qquad \qquad +\de_{i2}M_2^{ag}(\vec x)G_2(-\vec X) f^{gch} M_2^{rh}(\vec x+\vec X) [M_2^{-1}(x_1,v_2)M_1(x_1,v_2)M_1^{-1}(\vec v)]^{cb} \nn\\
&\qquad \qquad -[M_2(\vec x)M_2^{-1}(x_1,v_2)]^{ac} \de_{i2}M_2^{cg}(x_1,v_2)G_2((x_1,v_2)-\vec{x}-\vec X) f^{gdh} M_2^{rh}(\vec x+\vec X) \nn\\
&\qquad \qquad\qquad \times [M_2^{-1}(x_1,v_2)M_1(x_1,v_2)M_1^{-1}(\vec v)]^{db} \nn\\
&\qquad \qquad +[M_2(\vec x)M_2^{-1}(x_1,v_2)]^{ac} \de_{i1}M_1^{cg}(x_1,v_2)G_1((x_1,v_2)-\vec{x}-\vec X) f^{gdh} M_1^{rh}(\vec x+\vec X) [M_1^{-1}(\vec v)]^{db} \nn\\
&\qquad \qquad -[M_2(\vec x)M_2^{-1}(x_1,v_2)M_1(x_1,v_2)M_1^{-1}(\vec v)]^{ac}\nn\\
&\qquad \qquad\qquad \times  \de_{i1}M_1^{cg}(\vec v)G_1(\vec v-\vec{x}-\vec X) f^{gdh} M_1^{rh}(\vec x+\vec X) [M_1^{-1}(\vec v)]^{db} \Bigg] \frac{\de}{\de A_i^b(\vec v)} 
\nn
\\
&
=:\lim_{\nu\to\infty}\sum_{i=1}^8T_i\,.
\eE
In the third, fourth, seventh and and eighth term we can take the limit of $\nu\to\infty$ without problems. With $\Phi_{ar}(\vec x,\vec x) = \de^{ar}$ and after integrating the delta functions inside the Green's functions we find for these terms:
\bE{ll}
\lim_{\nu\to\infty}(&T_3+T_4+T_7+T_8) \nn\\
&={e\over4}\int_{x,v_2}\frac{\mu}{\sqrt{\pi}}\de_\mu(x_2-v_2)  \Bigg[[M_1(\vec x)M_1^{-1}(\vec x)]^{ac} M_2^{cg}(\vec x)\th(0) f^{gdh} M_2^{ah}(\vec x) [M_2^{-1}(x_1,v_2)]^{db} \nn\\
&\qquad \qquad -[M_1(\vec x)M_1^{-1}(\vec x)M_2(\vec x)M_2^{-1}(\vec v)]^{ac} \nn\\
&\qquad \qquad\qquad \times  M_2^{cg}(\vec v)\th(v_2-x_2) f^{gdh} M_2^{ah}(\vec x) [M_2^{-1}(x_1,v_2)]^{db} \Bigg] \frac{\de}{\de A_2^b(x_1,v_2)} \nn\\ 
&\quad + {e\over4}\int_{x,v_1} \de_\mu(x_1-v_1)\frac{\mu}{\sqrt{\pi}}\Bigg[[M_2(\vec x)M_2^{-1}(\vec x)]^{ac} M_1^{cg}(\vec x)\th(0) f^{gdh} M_1^{ah}(\vec x) [M_1^{-1}(v_1,x_2)]^{db} \nn\\
&\qquad \qquad -[M_2(\vec x)M_2^{-1}(\vec x)M_1(\vec x)M_1^{-1}(\vec v)]^{ac}\nn\\
&\qquad \qquad\qquad \times M_1^{cg}(\vec v)\th(v_1-x_1) f^{gdh} M_1^{ah}(\vec x) [M_1^{-1}(v_1,x_2)]^{db} \Bigg] \frac{\de}{\de A_1^b(v_1,x_2)} \\
&=0
\eE
All of these terms vanish under color contraction. We are thus left with
\bE{ll}
\half&\lim_{\nu\to\infty}\sum_i\int_{x,v,X}\de_\mu(\vec x,\vec v)\de_\nu(\vec X)\Phi_{ar}(\vec x,\vec x+\vec X)\left[\frac{\de}{\de A_i^r(\vec x+\vec X)}\Phi_{ab}(\vec x,\vec v)\right]\frac{\de}{\de A_i^b(\vec v)} \nn\\
&=\lim_{\nu\to\infty}(T_1+T_2+T_5+T_6) \nn\\
&={e\over4}\lim_{\nu\to\infty}\int_{X,x,v}\Phi_{ar}(\vec x,\vec x+\vec X)\de_\mu(\vec x,\vec v)\de_\nu(\vec X) \Big[\th(-X_1)-\th(v_1-x_1-X_1)\Big]\de(-X_2) \nn\\
&\qquad\qquad M_1^{ag}(\vec x)f^{gch} M_1^{rh}(\vec x+\vec X) [M_1^{-1}(v_1,x_2)M_2(v_1,x_2)M_2^{-1}(\vec v) ]^{cb} \frac{\de}{\de A_1^b(\vec v)} \nn\\
&\quad +{e\over4}\lim_{\nu\to\infty}\int_{X,x,v}\Phi_{ar}(\vec x,\vec x+\vec X)\de_\mu(\vec x,\vec v)\de_\nu(\vec X) \de(-X_1)\Big[\th(-X_2)-\th(v_2-x_2-X_2)\Big]  \nn\\
&\qquad\qquad M_2^{ag}(\vec x) f^{gch} M_2^{rh}(\vec x +\vec X) [M_2^{-1}(x_1,v_2)M_1(x_1,v_2)M_1^{-1}(\vec v)]^{cb}  \frac{\de}{\de A_2^b(\vec v)} \\
&={e\over4}\lim_{\nu\to\infty}\int_{X_1,x,v}\Phi_{ar}(\vec x,\vec x+\vec X)|_{X_2=0} \de_\mu(\vec x,\vec v)\de_\nu(X_1)\frac{\nu}{\sqrt{\pi}} \Big[\th(-X_1)-\th(v_1-x_1-X_1)\Big] \nn\\
&\qquad\qquad M_1^{ag}(\vec x)f^{gch} M_1^{rh}(x_1+X_1,x_2) [M_1^{-1}(v_1,x_2)M_2(v_1,x_2)M_2^{-1}(\vec v) ]^{cb} \frac{\de}{\de A_1^b(\vec v)} \nn\\
&\quad +{e\over4}\lim_{\nu\to\infty}\int_{X_2,x,v}\Phi_{ar}(\vec x,\vec x+\vec X)|_{X_1=0} \de_\mu(\vec x,\vec v)\de_\nu(X_2) \frac{\nu}{\sqrt{\pi}} \Big[\th(-X_2)-\th(v_2-x_2-X_2)\Big]  \nn\\
&\qquad\qquad M_2^{ag}(\vec x) f^{gch} M_2^{rh}(x_1,x_2+X_2) [M_2^{-1}(x_1,v_2)M_1(x_1,v_2)M_1^{-1}(\vec v)]^{cb}  \frac{\de}{\de A_2^b(\vec v)} \,.
\eE
With Eq.~(\ref{Phi}):
\be
\Phi^{ab}(u,v)= {1\over 2} (M_1(u)M_1^{-1}(v_1,u_2)M_2(v_1,u_2)M_2^{-1}(v)+M_2(u)M_2^{-1}(u_1,v_2)M_1(u_1,v_2)M_1^{-1}(v))^{ab} 
\ee
this is
\bE{ll}
&={e\over4}\lim_{\nu\to\infty}\int_{X_1,x,v}(M_1(\vec x)M_1^{-1}(x_1+X_1,x_2))^{ar}  \de_\mu(\vec x,\vec v)\de_\nu(X_1)\frac{\nu}{\sqrt{\pi}} \Big[\th(-X_1)-\th(v_1-x_1-X_1)\Big] \nn\\
&\qquad\qquad M_1^{ag}(\vec x)f^{gch} M_1^{rh}(x_1+X_1,x_2) [M_1^{-1}(v_1,x_2)M_2(v_1,x_2)M_2^{-1}(\vec v) ]^{cb} \frac{\de}{\de A_1^b(\vec v)} \nn\\
&\quad +{e\over4}\lim_{\nu\to\infty}\int_{X_2,x,v} (M_2(\vec x)M_2^{-1}(x_1,x_2+X_2))^{ar}\de_\mu(\vec x,\vec v)\de_\nu(X_2) \frac{\nu}{\sqrt{\pi}} \Big[\th(-X_2)-\th(v_2-x_2-X_2)\Big]  \nn\\
&\qquad\qquad M_2^{ag}(\vec x) f^{gch} M_2^{rh}(x_1,x_2+X_2) [M_2^{-1}(x_1,v_2)M_1(x_1,v_2)M_1^{-1}(\vec v)]^{cb}  \frac{\de}{\de A_2^b(\vec v)} \\
&={e\over4}\lim_{\nu\to\infty}\int_{X_1,x,v}\de_\mu(\vec x,\vec v)\de_\nu(X_1)\frac{\nu}{\sqrt{\pi}} \Big[\th(-X_1)-\th(v_1-x_1-X_1)\Big] \nn\\
&\qquad\qquad \de^{gh}f^{gch} [M_1^{-1}(v_1,x_2)M_2(v_1,x_2)M_2^{-1}(\vec v) ]^{cb} \frac{\de}{\de A_1^b(\vec v)} \nn\\
&\quad +{e\over4}\lim_{\nu\to\infty}\int_{X_2,x,v} \de_\mu(\vec x,\vec v)\de_\nu(X_2) \frac{\nu}{\sqrt{\pi}} \Big[\th(-X_2)-\th(v_2-x_2-X_2)\Big]  \nn\\
&\qquad\qquad \de^{gh}f^{gch} [M_2^{-1}(x_1,v_2)M_1(x_1,v_2)M_1^{-1}(\vec v)]^{cb}  \frac{\de}{\de A_2^b(\vec v)} \\
&=0
\eE
Again, these terms vanish under color contraction. Hence we conclude, that
\be
\mathcal{T}_{reg}=-{1\over2}\int_{x,v} \delta_\mu(\vec x, \vec v)  \frac{\delta}{\delta A_i^a(\vec x)}  \Phi_{ab}(\vec x,\vec v) \frac{\delta}{\delta A_i^b(\vec v)} =-{1\over2}\int_{x,v} \delta_\mu(\vec x, \vec v) \Phi_{ab}(\vec x,\vec v) \frac{\delta}{\delta A_i^a(\vec x)}  \frac{\delta}{\delta A_i^b(\vec v)}\,.
\label{Tregeq}
\ee
to any order in perturbation theory. This confirms that Eq. (\ref{Treg}) is Hermitian.
Finally, as a check, we have also performed the above computation, using the explicit form of the string, to ${\cal O}(e^2)$. 

\section{Useful equalities for Sec.~\ref{sec:KKN}}
\label{sec:useful}
In this Appendix we compile a series of useful equalities and computations that we have used in Sec.~\ref{sec:KKN}. 

Inverting equations (\ref{AbarA}) yields (for a more compact expression see Eq.~(5) of \cite{Karabali:1996iu})
\bE{rCl}
\label{Mexp}
M(\vec x)&=&1-e\frac{4}{\vec \nabla^2}(\bar \partial A)+e^2\frac{4}{\vec \nabla^2}\bar \partial A\frac{4}{\vec \nabla^2}\bar \partial A
+O(e^3) \\
&=&1-e\int_y G(\bar x;\bar y) A(\vec y) + e^2\int_{y,z}G(\bar x;\bar z)A(\vec z)G(\bar z;\bar y)A(\vec y)+ O(e^3) \label{MofA} \,, \\
M^{\dagger}(\vec x) &=& 1+e\frac{4}{\vec \nabla^2}( \partial \bar A)+e^2\frac{4}{\vec \nabla^2} \partial \left(\frac{4}{\vec \nabla^2} \partial\bar A\right) \bar A + O(e^3) \\
&=&1+e\int_y \bar{G}(x;y) \bar A(\vec y) + e^2\int_{y,z}\bar G(x;z) \bar G(z;y)\bar A(\vec y) \bar A(\vec z)+ O(e^3)  \label{MdaggerofAbar}
\,,
\eE
with the Green functions:
\bea
\bar G(x;y) &\equiv& \bar G(x-y) = \frac{1}{\bar \partial_x}\delta^{(2)}(\vec x -\vec y)=-i\int \frac{d^2k}{(2\pi)^2}e^{i\vec k \cdot(\vec x -\vec y)}\frac{1}{\bar k}=\frac{1}{\pi}\frac{(\bar x-\bar y)}{(x-y)(\bar x-\bar y)+\epsilon^2} \label{Gbar}
\,,\quad 
\\
 G(\bar x; \bar y) &\equiv& G(\bar x-\bar y) = \frac{1}{\partial_x}\delta^{(2)}(\vec x -\vec y)=-i\int \frac{d^2k}{(2\pi)^2}e^{i\vec k \cdot(\vec x -\vec y)}\frac{1}{k}=\frac{1}{\pi}\frac{(x-y)}{(x-y)(\bar x-\bar y)+\epsilon^2}\label{G}
\,.\quad 
\eea
These are the holomorphic and anti-holomorphic analogues of Eqs.~(\ref{M_iofA_i}-\ref{G_i}).\\

We also need ($T_F=1/2$)
\bea
\left(M^{\dagger}\right)^{ac}=\frac{1}{T_F}Tr[T^aM^{\dagger}T^cM^{\dagger-1}]
\,,
\eea
and the analogue for $M^{ac}$ (note that $M^{-1}_{ac}=M_{ca}$). With this definition one can easily check the following identity
\be
M^{\dagger}_{cg}f^{gbh}M^{\dagger-1}_{hd} = -f^{cdf}M^{\dagger-1}_{bf} \,. \label{Mf}
\ee 

Some useful relations are:
\bea
D&=&\p+eA=M\p M^{-1}\,,\qquad \bar{D}=\bar\p+e\bar A=M^{\dagger-1} \bar\partial M^{\dagger} \,,\\
\left(\frac{1}{\bar D}\right)^{de}_{yx} \label{DbarInv}
&=&
\bar G(y-x)\left[M^{\dagger-1}(\vec y)M^{\dagger}(\vec x)\right]_{de}\,, \\
\frac{\delta M^{\dagger}_{cd}(\vec y)}{\delta \bar A^b(\vec x)}&=&
e\left(\frac{1}{\bar D}\right)^{de}_{yx}(-f_{ebh})M_{hc}^{\dagger-1}(\vec x)=
e\left(\frac{1}{\bar D}\right)^{eb}_{yx} f_{edh} M_{hc}^{\dagger-1}(\vec y)
\,. \label{MdaggerOverAbar}
\eea
\bea
\frac{\delta J^c(\vec y)}{\delta A^b(\vec x)}&=&2i M^{\dagger}_{cb}(\vec y)\delta(\vec y-\vec x)
\,,
\\
\frac{\delta J^c(\vec y)}{\delta \bar A^b(\vec x)}&=&2\left[i\frac{\delta M^{\dagger}_{cd}(\vec y)}{\delta \bar A^b(\vec x)}A_d(\vec y)
+\frac{1}{e}\frac{\delta }{\delta \bar A^b(\vec x)}\left((\partial M^{\dagger}(\vec y))M^{\dagger-1}(\vec y)\right)_c
\right]
\,.
\eea

With Eqs.~(\ref{H}), (\ref{J}), (\ref{DbarInv}) and (\ref{MdaggerOverAbar}) we find
\bE{rCl}
M^\dagger_{dh}(\vec z) D_z^{he}\left(\bar{D}^{-1}\right)^{ea}_{zx} &=& M^\dagger_{dh}(\vec z) \left(M(\vec z)\p_z M^{-1}(\vec z)\right)^{he} \left(M^{\dagger-1}(\vec z)\bar{G}(z-x) M^\dagger(\vec x)\right)^{ea} \\
&=& \left(H(\vec z)\p_z H^{-1}(\vec z)\bar{G}(z-x) M^\dagger(\vec x)\right)^{da} \\
&=& \left(\p_z \bar{G}(z-x)\right)M^\dagger_{da}(\vec x) -\frac{e}{2}\bar{G}(z-x) \left(J(\vec z)M^\dagger(\vec x)\right)^{da} \\
&=& \left(\p_z \bar{G}(z-x)\right)M^\dagger_{da}(\vec x) + \frac{ie}{2}\bar{G}(z-x) f^{edf} J^e(\vec z)M^{\dagger}_{fa}(\vec x)\,,
\eE
or, more compact and for further reference:
\bE{rCl}
\label{JOverAbar}
\frac{\delta J^d(\vec z)}{\delta \bar{A}^a(\vec x)} 
&=& -2i\left(\mathcal{D}_z\bar{G}(z-x)M^{\dagger}(\vec x)\right)^{da}\,,\\
\mathcal{D}^{mn} &=& \p_z \delta^{mn} + \frac{ie}{2} f^{mnc} J^c(\vec z)\,.
\eE

\section{Computations of the vanishing terms}

\subsection{Order $e$ correction to the gauge field Hamiltonian}
\label{subsec:append4}

\bE{l}
-{1\over2}\int_{u,v}\delta_\mu(\vec u,\vec v) \Phi_{ab}^{(1)}(\vec u,\vec v) \frac{\delta F_{GL}^{(0)}}{\delta  A_i^a(\vec u)} \frac{\delta F_{GL}^{(0)}}{\delta A_i^b(\vec v)} \nn\\
= - {1\over 8\pi^2} \int_{u,v,y,w}\p_{u_i}\p_{v_i}  \Big( \delta_\mu(\vec u,\vec v) \Phi_{ab}^{(1)}(\vec u,\vec v)  \Big)  \frac{1}{|\vec{u}-\vec{y}|}\frac{1}{|\vec{v}-\vec{w}|}  (\vec{\nabla}\times\vec{A}^a(\vec{y}))(\vec{\nabla}\times\vec{A}^b(\vec{w}))  \\
= - {e\over 4\pi^2}f^{abc} \int_{U,v,x,y,w}   \delta_\mu(\vec U)\nn\\
\quad \Bigg(   (G_1(\vec U+\vec v-\vec y)-G_1(v_1-y_1,U_2+v_2-y_2) +G_1(U_1+v_1-y_1,v_2-y_2)-G_1(\vec v-\vec y)) A_1^c(\vec y)  \nn\\
\qquad  +  (G_2(v_1-y_1,U_2+v_2-y_2)-G_2(\vec v-\vec y)+G_2(\vec U+\vec v-\vec y)-G_2(U_1+v_1-y_1,v_2-y_2)) A_2^c(\vec y) \Bigg) \nn\\
\quad \frac{\mu^2 -\mu^4|\vec{U}|^2}{|\vec{U}+\vec{v}-\vec{x}||\vec{v}-\vec{w}|}  (\vec{\nabla}\times\vec{A}^a(\vec{x}))(\vec{\nabla}\times\vec{A}^b(\vec{w})) \nn\\
\quad -{e\over 4\pi^2}f^{abc} \int_{U,v,x,w}   \delta_\mu(U)  \frac{\mu^2}{|\vec{U}+\vec{v}-\vec{x}||\vec{v}-\vec{w}|}  (\vec{\nabla}\times\vec{A}^a(\vec{x}))(\vec{\nabla}\times\vec{A}^b(\vec{w}))\nn\\
 \quad \Bigg\{ U_1 (A_1^c(v_1,U_2+v_2)+A_1^c(\vec v)) + U_2 (A_2^c(\vec v)+A_2^c(U_1+v_1,v_2)) \Bigg\} \nn\\
\quad +{e\over 4\pi^2}f^{abc} \int_{U,v,x,y,w}   \delta_\mu(U)  \frac{\mu^2}{|\vec{U}+\vec{v}-\vec{x}||\vec{v}-\vec{w}|}  (\vec{\nabla}\times\vec{A}^a(\vec{x}))(\vec{\nabla}\times\vec{A}^b(\vec{w}))\nn\\ \quad \Bigg\{U_1 (G_2(v_1,U_2+v_1;\vec y)-G_2(\vec v;\vec y)) \p_1A_2^c(\vec y)  + U_2 (G_1(U_1+v_1,v_2;\vec y)-G_1(\vec v;\vec y)) \p_2 A_1^c(\vec y)  \Bigg\} 
\eE
Except for $\delta_\mu(U)$, we Taylor expand this expression in powers of $U$. The first integral up to 4th order, the other two up to 2nd order.
\bE{l}
= - {e\over 4\pi^2}f^{abc} \int_{v,x,w} \Bigg(\frac{1}{|\vec{v}-\vec{x}|^2} (\vec{v}-\vec{x})\cdot\vec{A}^c(\vec{v}) +\frac{1}{2}\nabla\cdot\vec{A}^c(\vec{v})   \Bigg) \frac{1}{|\vec{v}-\vec{x}||\vec{v}-\vec{w}|}  (\vec{\nabla}\times\vec{A}^a(\vec{x}))(\vec{\nabla}\times\vec{A}^b(\vec{w})) \nn\\
\quad -{e\over 4\pi^2}f^{abc} \int_{v,x,w}  \frac{1}{|\vec{v}-\vec{x}||\vec{v}-\vec{w}|}  (\vec{\nabla}\times\vec{A}^a(\vec{x}))(\vec{\nabla}\times\vec{A}^b(\vec{w}))  \Bigg\{ -\frac{1}{|\vec{v}-\vec{x}|^2} (\vec{v}-\vec{x})\cdot\vec{A}^c(\vec{v}) \Bigg\} \nn\\
\quad +{e\over 4\pi^2}f^{abc} \int_{v,x,y,w}    \frac{\mu^2}{|\vec{v}-\vec{x}||\vec{v}-\vec{w}|}  (\vec{\nabla}\times\vec{A}^a(\vec{x}))(\vec{\nabla}\times\vec{A}^b(\vec{w})) \Bigg\{ 0 \Bigg\} +\mathcal{O}(\mu^{-2})
\eE
\bE{l}
 = - {e\over 8\pi^2}f^{abc} \int_{v,x,w}  \frac{1}{|\vec{v}-\vec{x}||\vec{v}-\vec{w}|}  (\vec{\nabla}\times\vec{A}^a(\vec{x}))(\vec{\nabla}\times\vec{A}^b(\vec{w})) (\nabla\cdot\vec{A}^c(\vec{v})) + \mathcal{O}(\mu^{-2}) \\
=\mathcal{O}(\mu^{-2})
\eE
The $\mathcal{O}(\mu^0)$ term vanishes under combined interchange of $\{x\leftrightarrow w,a\leftrightarrow b\}$.

\subsection{Order $e^2A^4$ correction to the gauge field Hamiltonian}
\label{subsec:append5}
Vanishing of the first term:
\bE{l}
-{1\over2}\int_{u,v}\delta_\mu(\vec u,\vec v) \Phi_{ab}^{(2)}(\vec u,\vec v) \frac{\delta F_{GL}^{(0)}}{\delta  A_i^a(\vec u)} \frac{\delta F_{GL}^{(0)}}{\delta A_i^b(\vec v)} \nn\\
= - {1\over 8\pi^2} \int_{u,v,r,w}\p_{u_i}\p_{v_i}  \Big( \delta_\mu(\vec u,\vec v) \Phi_{ab}^{(2)}(\vec u,\vec v)  \Big)  \frac{1}{|\vec{u}-\vec{r}|}\frac{1}{|\vec{v}-\vec{w}|}  (\vec{\nabla}\times\vec{A}^a(\vec{r}))(\vec{\nabla}\times\vec{A}^b(\vec{w}))  \\
= - {1\over 4\pi^2}  f^{adc}f^{dbe} \int_{u,v,r,w,y,z} \delta_\mu(\vec u,\vec v)  \frac{\mu^2-\mu^4(\vec{u}-\vec{v})^2}{|\vec{u}-\vec{r}||\vec{v}-\vec{w}|}  (\vec{\nabla}\times\vec{A}^a(\vec{r}))(\vec{\nabla}\times\vec{A}^b(\vec{w})) \nn\\
\qquad\Bigg\{ \Big( (G_1(\vec u;\vec z)-G_1(v_1,u_2;\vec z))(G_1(\vec z;\vec y)-G_1(v_1,u_2;\vec y))\nn\\
\qquad \qquad + (G_1(u_1,v_2;\vec z)-G_1(\vec v;\vec z))(G_1(\vec z;\vec y)-G_1(\vec v;\vec y))\Big)A_1^c(\vec z)A_1^e(\vec y)  \nn\\
 \qquad+  \Big( (G_2(v_1,u_2;\vec z)-G_2(\vec v;\vec z))(G_2(\vec z;\vec y)-G_2(\vec v;\vec y))\nn\\
 \qquad\qquad+ (G_2(\vec u;\vec z)-G_2(u_1,v_2;\vec z))(G_2(\vec z;\vec y)-G_2(u_1,v_2;\vec y))\Big)A_2^c(\vec z)A_2^e(\vec y)  \nn\\
\qquad +   (G_1(\vec u;\vec y)-G_1(v_1,u_2;\vec y))(G_2(v_1,u_2;\vec z)-G_2(\vec v;\vec z))A_1^c(\vec y) A_2^e(\vec z) \nn\\
 \qquad+  (G_2(\vec u;\vec z)-G_2(u_1,v_2;\vec z))(G_1(u_1,v_2;\vec y)-G_1(\vec v;\vec y))A_2^c(\vec z)A_1^e(\vec y) \Bigg\} \nn\\
 + {1\over 4\pi^2} f^{adc}f^{dbe} \int_{u,v,r,w,y,z}  \frac{\mu^2}{|\vec{u}-\vec{r}| |\vec{v}-\vec{w}|}  (\vec{\nabla}\times\vec{A}^a(\vec{r}))(\vec{\nabla}\times\vec{A}^b(\vec{w}))  \delta_\mu(\vec u,\vec v) \nn\\
\qquad \Bigg\{(u_1-v_1)\Bigg( (G_2(v_1,u_2;\vec z)-G_2(\vec v;\vec z))(G_2(\vec z;\vec y)-G_2(\vec v;\vec y)) (\p_1A_2^c(\vec z)A_2^e(\vec y) +A_2^c(\vec z)\p_1A_2^e(\vec y)  ) \nn\\
\qquad +  (G_1(\vec u;\vec y)-G_1(v_1,u_2;\vec y)) (G_2(v_1,u_2;\vec z)-G_2(\vec v;\vec z)) A_1^c(\vec y)  \p_1 A_2^e(\vec z)  \Bigg) \nn\\
\qquad +(u_2-v_2) \Bigg(  (G_1(u_1,v_2;\vec z)-G_1(\vec v;\vec z))(G_1(\vec z;\vec y)-G_1(\vec v;\vec y)) (\p_2A_1^c(\vec z)A_1^e(\vec y) + A_1^c(\vec z) \p_2A_1^e(\vec y) ) \nn\\
 \qquad+  (G_2(\vec u;\vec z)-G_2(u_1,v_2;\vec z))(G_1(u_1,v_2;\vec y)-G_1(\vec v;\vec y)) A_2^c(\vec z)\p_2A_1^e(\vec y) \Bigg) \Bigg\} \nn\\ 
- {1\over 16\pi^2} f^{adc}f^{dbe} \int_{u,v,r,w,z}  \frac{1}{|\vec{u}-\vec{r}| |\vec{v}-\vec{w}|}  (\vec{\nabla}\times\vec{A}^a(\vec{r}))(\vec{\nabla}\times\vec{A}^b(\vec{w}))  \delta_\mu(\vec u,\vec v) \nn\\
\qquad \Bigg\{  (G_2(v_1,u_2;\vec z)-G_2(\vec v;\vec z)) A_1^c(\vec u)  \p_1A_2^e(\vec z)  -(G_2(\vec u;\vec z)-G_2(u_1,v_2;\vec z))  \p_1 A_2^c(\vec z)A_1^e(\vec v) \nn\\
\qquad + (G_1(u_1,v_2;\vec z)-G_1(\vec v;\vec z))A_2^c(\vec u)\p_2A_1^e(\vec z)  - (G_1(\vec u;\vec z)-G_1(v_1,u_2;\vec z))  \p_2A_1^c(\vec z) A_2^e(\vec v) \Bigg\} \nn\\
- {1\over 4\pi^2} f^{adc}f^{dbe} \int_{u,v,r,w,z}  \frac{\mu^2}{|\vec{u}-\vec{r}||\vec{v}-\vec{w}|}  (\vec{\nabla}\times\vec{A}^a(\vec{r}))(\vec{\nabla}\times\vec{A}^b(\vec{w}))  \delta_\mu(\vec u,\vec v) \nn\\
\qquad \Bigg\{ (u_1-v_1)\Bigg(  (G_1(\vec u;\vec z)-G_1(v_1,u_2;\vec z))A_1^e(v_1,u_2) A_1^c(\vec z) + (G_1(u_1,v_2;\vec z)-G_1(\vec v;\vec z))A_1^e(\vec v) A_1^c(\vec z)  \nn\\
\qquad  +(G_2(v_1,u_2;\vec z)-G_2(\vec v;\vec z))  A_1^c(v_1,u_2) A_2^e(\vec z) + (G_2(\vec u;\vec z)-G_2(u_1,v_2;\vec z))A_2^c(\vec z)A_1^e(\vec v) \Bigg) \nn\\
\qquad +(u_2-v_2) \Bigg( (G_2(\vec u;\vec z)-G_2(u_1,v_2;\vec z)) A_2^e(u_1,v_2)A_2^c(\vec z) + (G_2(v_1,u_2;\vec z)-G_2(\vec v;\vec z))A_2^e(\vec v) A_2^c(\vec z)  \nn\\
\qquad +   (G_1(u_1,v_2;\vec z)-G_1(\vec v;\vec z)) A_2^c(u_1,v_2) A_1^e(\vec z) +  (G_1(\vec u;\vec z)-G_1(v_1,u_2;\vec z))A_2^e(\vec v) A_1^c(\vec z)  \Bigg) \Bigg\} \nn\\
+ {1\over 8\pi^2} f^{adc}f^{dbe} \int_{u,v,r,w}  \frac{1}{|\vec{u}-\vec{r}|}\frac{1}{|\vec{v}-\vec{w}|}  (\vec{\nabla}\times\vec{A}^a(\vec{r}))(\vec{\nabla}\times\vec{A}^b(\vec{w}))  \delta_\mu(\vec u,\vec v) \nn\\
\qquad \Bigg\{ \Big( A_1^c(\vec u)A_1^e(v_1,u_2) +A_2^c(\vec u) A_2^e(u_1,v_2)\Big)  \Bigg\}  
\eE
\bE{l}
= - {1\over 4\pi^2}  f^{adc}f^{dbe} \int_{r,w,y,z} \frac{ -2\delta(y-z)}{4|\vec{z}-\vec{r}||\vec{z}-\vec{w}|}  (\vec{\nabla}\times\vec{A}^a(\vec{r}))(\vec{\nabla}\times\vec{A}^b(\vec{w}))  \Bigg\{A_1^c(\vec z)A_1^e(\vec y) +A_2^c(\vec z)A_2^e(\vec y)   \Bigg\}  \nn\\
\quad - {1\over 4\pi^2} f^{adc}f^{dbe} \int_{r,w,z}  \frac{2}{2|\vec{z}-\vec{r}||\vec{z}-\vec{w}|}  (\vec{\nabla}\times\vec{A}^a(\vec{r}))(\vec{\nabla}\times\vec{A}^b(\vec{w}))    \Bigg\{ A_1^e(\vec z) A_1^c(\vec z) +A_2^e(\vec z) A_2^c(\vec z)   \Bigg\} \nn\\
\quad + {1\over 8\pi^2} f^{adc}f^{dbe} \int_{u,r,w}  \frac{1}{|\vec{u}-\vec{r}|}\frac{1}{|\vec{u}-\vec{w}|}  (\vec{\nabla}\times\vec{A}^a(\vec{r}))(\vec{\nabla}\times\vec{A}^b(\vec{w}))    \Bigg\{ A_1^c(\vec u)A_1^e(\vec u) +A_2^c(\vec u) A_2^e(\vec u)  \Bigg\} \nn\\
\quad +\mathcal{O}(\mu^{-2})  \nn\\
=0+\mathcal{O}(\mu^{-2}) 
\eE
This vanishes for $\mu\to\infty$.\\

Vanishing of the second term:

\bE{l}
 -\int_{u,v} \delta_\mu(\vec u,\vec v)  \Phi_{ab}^{(1)}(u,v)  \frac{\delta F_{GL}^{(0)}}{\delta A_i^a(\vec u)} \frac{\delta F_{GL}^{(1)}}{\delta A_i^b(\vec v)}  \nn\\ 
 =  - {i\over2} f^{a_1a_2b} f^{abc}\int_{\slashed{k_1},\slashed{k_2},\slashed{q},\slashed{p}} \int_{u,v,y} \delta_\mu(\vec u,\vec v)   \Bigg\{ (G_1(\vec u;\vec y)-G_1(v_1,u_2;\vec y)+G_1(u_1,v_2;\vec y)-G_1(\vec v;\vec y)) A_1^c(\vec y)  \nn\\
\qquad + (G_2(v_1,u_2;\vec y)-G_2(\vec v;\vec y)+G_2(\vec u;\vec y)-G_2(u_1,v_2;\vec y)) A_2^c(\vec y)  \Bigg\}\nn\\
 \qquad   e^{-i\vec{q}\cdot\vec{v}} e^{-i\vec{p}\cdot\vec{u}} \frac{1}{|\vec{p}|}(\vec{p}\times\vec{A}^a(-\vec{p}))  \frac{\slashed{\delta}\left(\vec{k}_1+\vec{k}_2+\vec{q}\right)}{|\vec{k}_1|+|\vec{k}_2|+|\vec{q}|} \Bigg\{ \frac{1}{2}  \vec{p}\cdot\vec{q}  (\vec{A}^{a_1}(\vec{k}_1)\times\vec{A}^{a_2}(\vec{k}_2))\nn\\
\qquad + (\vec{k}_1\times\vec{A}^{a_1}(\vec{k}_1)) \vec{p}\cdot\vec{A}^{a_2} (\vec{k}_2) -\frac{\vec{p}\times\vec{q}}{|\vec{q}||\vec{k}_2|}   (\vec{k}_2\times\vec{A}^{a_1}(\vec{k}_1)) (\vec{k}_2\times\vec{A}^{a_2}(\vec{k}_2)) \nn\\
\qquad +\frac{\vec{p}\cdot\vec{k}_2}{|\vec{k}_1||\vec{k}_2|}  (\vec{k}_1\cdot\vec{A}^{a_1}(\vec{k}_1))  (\vec{k}_2\times\vec{A}^{a_2}(\vec{k}_2))  -\frac{\vec{p}\cdot\vec{q}}{|\vec{k}_1||\vec{q}|} (\vec{k}_1\cdot\vec{A}^{a_1}(\vec{k}_1)) (\vec{q}\times\vec{A}^{a_2}(\vec{k}_2))  \Bigg\}   \\
 = {1\over2} f^{a_1a_2b} f^{abc}\int_{\slashed{k_1},\slashed{k_2},\slashed{p},\slashed{r}} \sd(p-k_1-k_2-r)  \Bigg\{ \left(e^{-\frac{p_1^2}{4 \mu ^2}}-e^{-\frac{(k_{1,1}+k_{2,1})^2}{4 \mu ^2}}\right) \left(e^{-\frac{p_2^2}{4 \mu ^2}}+e^{-\frac{(k_{1,2}+k_{2,2})^2}{4 \mu ^2}}\right)\frac{1}{r_1} A_1^c(r)  \nn\\
\qquad + \left(e^{-\frac{p_1^2}{4 \mu ^2}}+e^{-\frac{(k_{1,1}+k_{2,1})^2}{4 \mu ^2}}\right) \left(e^{-\frac{p_2^2}{4 \mu ^2}}-e^{-\frac{(k_{1,2}+k_{2,2})^2}{4 \mu ^2}}\right)\frac{1}{r_2}  A_2^c(r)  \Bigg\}\nn\\
 \qquad    \frac{1}{|\vec{p}|}(\vec{p}\times\vec{A}^a(-\vec{p}))  \frac{1}{|\vec{k}_1|+|\vec{k}_2|+|\vec{k}_1+\vec{k}_2|} \Bigg\{- \frac{1}{2}  \vec{p}\cdot(\vec{k}_1+\vec{k}_2)  (\vec{A}^{a_1}(\vec{k}_1)\times\vec{A}^{a_2}(\vec{k}_2))\nn\\
\qquad + (\vec{k}_1\times\vec{A}^{a_1}(\vec{k}_1)) \vec{p}\cdot\vec{A}^{a_2} (\vec{k}_2) +\frac{\vec{p}\times(\vec{k}_1+\vec{k}_2)}{|\vec{k}_1+\vec{k}_2||\vec{k}_2|}   (\vec{k}_2\times\vec{A}^{a_1}(\vec{k}_1)) (\vec{k}_2\times\vec{A}^{a_2}(\vec{k}_2)) \nn\\
\qquad +\frac{\vec{p}\cdot\vec{k}_2}{|\vec{k}_1||\vec{k}_2|}  (\vec{k}_1\cdot\vec{A}^{a_1}(\vec{k}_1))  (\vec{k}_2\times\vec{A}^{a_2}(\vec{k}_2))  -\frac{\vec{p}\cdot(\vec{k}_1+\vec{k}_2)}{|\vec{k}_1||\vec{k}_1+\vec{k}_2|} (\vec{k}_1\cdot\vec{A}^{a_1}(\vec{k}_1)) (\vec{k}_1+\vec{k}_2)\times\vec{A}^{a_2}(\vec{k}_2)  \Bigg\}   
 \eE

This vanishes for $\mu\to\infty$.

\subsection{Order $e$ correction to the gauge invariant Hamiltonian}
\label{subsec:append1}
Vanishing of the second term:
\bE{rl}
&\int_{x,v,y}\tilde{\Omega}^{(1)}_{ab}(\vec x,\vec v,\vec y)\frac{\delta F_{GI}^{(0)}}{\delta J^a(\vec x)} \frac{\delta F_{GI}^{(0)}}{\delta J^b(\vec y)}  \nn\\
\propto & \int_{x,v,y,w,z,r,s} \delta_\mu(\vec x, \vec v) (x-v) \left(\p_y \bar{G}(y-v)\right)f^{abe}J^e(\vec v) \frac{1}{|\vec{w}-\vec{z}|}\bar{\p}_w\de(\vec w-\vec x)J^a(\vec z) \frac{1}{|\vec{r}-\vec{s}|}\bar{\p}_r\de(\vec r-\vec y)J^b(\vec s) \qquad \\
\propto & \int_{x,v,y,w,z,r,s} \delta_\mu(\vec x, \vec v) \mu^2 (x-v)^2 \left(\p_y \de(\vec y-\vec v)\right)f^{abe}J^e(\vec v) \frac{1}{|\vec{w}-\vec{z}|}\de(\vec w-\vec x)J^a(\vec z) \frac{1}{|\vec{r}-\vec{s}|}\de(\vec r-\vec y)J^b(\vec s) \qquad \\
\propto & \int_{x,v,z,s} \delta_\mu(\vec x, \vec v) \mu^2 (x-v)^2 f^{abe}J^e(\vec v) \frac{1}{|\vec{x}-\vec{z}|} J^a(\vec z) \p_v\frac{1}{|\vec{v}-\vec{s}|}J^b(\vec s) \qquad
\eE 
Expanding $ \frac{1}{|\vec{x}-\vec{z}|}$ around $\vec{x}=\vec{v}$, we obtain
\bE{rl}
\propto & \int_{x,v,z,s} \delta_\mu(\vec x, \vec v) \mu^2 (x-v)^2 f^{abe}J^e(\vec v) J^a(\vec z) J^b(\vec s)\left( \frac{1}{|\vec{v}-\vec{z}|} \p_v\frac{1}{|\vec{v}-\vec{s}|} + (\vec{x}-\vec{v})\cdot\nabla_v \frac{1}{|\vec{v}-\vec{z}|} \p_v\frac{1}{|\vec{v}-\vec{s}|} + \ldots\right) \nn\\
=& O(\mu^{-2})
\eE 
Integration over $x$ vanishes for the first two orders (note that $(x-v)^2$ is only the holomorphic component), while the next order is already $O(\mu^{-2})$.

\subsection{Order $e^2J^4$ corrections to the gauge invariant Hamiltonian}
\label{subsec:append2}
Vanishing of the first term:
\bE{rl}
&\int_{x,v,y}\tilde{\Omega}^{(1)}_{ab}(\vec x,\vec v,\vec y)\frac{\delta F_{GI}^{(0)}}{\delta J^a(\vec x)} \frac{\delta F_{GI}^{(1)}}{\delta J^b(\vec y)}  \nn\\
\propto & f^{abe} \int_{x,v,y}\int_{\slashed{k_1},\slashed{k_2},\slashed{l},\slashed{p},\slashed{q}} \delta_\mu(\vec x, \vec v) (x-v) \frac{q}{\bar{q}} e^{i\vec{q}\cdot(\vec{y}-\vec{v})} J^e(\vec l) e^{i\vec{l}\cdot\vec{v}} \left(e^{-i\vec{p}\cdot\vec{x}} + e^{i\vec{p}\cdot\vec{x}} \right) \frac{\bar{p}^2}{|\vec p|} J^a(\vec{p}) \nn\\
 & e^{i(\vec{k}_1+\vec{k}_2)\cdot\vec{y}} f^{a_1 a_2 b}  g^{(3)}(\vec k_1,\vec k_2,-\vec k_1-\vec k_2) J^{a_1}(\vec k_1)J^{a_2}(\vec k_2) \\
\propto & \frac{1}{\mu^2} f^{abe} \int_{\slashed{k_1},\slashed{k_2},\slashed{l},}e^{-\frac{(\vec{k}_1+\vec{k}_2+\vec{l})^2}{4\mu^2}}  \frac{k_1+k_2}{\bar{k}_1+\bar{k}_2}  J^e(\vec l)  \frac{(\bar{k}_1+\bar{k}_2+\bar{l})^3}{|\vec{k}_1+\vec{k}_2+\vec{l}|} J^a(\vec{k}_1+\vec{k}_2+\vec{l}) \nn\\
 & f^{a_1 a_2 b}  g^{(3)}(\vec k_1,\vec k_2,-\vec k_1-\vec k_2) J^{a_1}(\vec k_1)J^{a_2}(\vec k_2) 
\eE 
This vanishes for $\mu\to\infty$

Vanishing of the second term:
\bE{rl}
&\int_{x,v,y}\tilde{\Omega}^{(2)}_{ab}(\vec x,\vec v,\vec y)\frac{\delta F_{GI}^{(0)}}{\delta J^a(\vec x)} \frac{\delta F_{GI}^{(0)}}{\delta J^b(\vec y)} \nn\\
\propto & \int_{x,v,y}\int_{\slashed{k},\slashed{l},\slashed{p},\slashed{q},\slashed{r}} \delta_\mu(\vec x, \vec v) f^{bec}f^{ead} \frac{1}{\bar{q}} e^{i\vec{q}\cdot(\vec{y}-\vec{v})} \Big( (x-v)^2 q e^{i\vec{l}\cdot\vec{v}} -2i(x-v)e^{i\vec{l}\cdot\vec{y}} \Big)J^c(\vec l) J^d(\vec r) e^{i\vec{r}\cdot\vec{v}} \nn\\
&  \left(e^{-i\vec{k}\cdot\vec{x}} + e^{i\vec{k}\cdot\vec{x}} \right) \left(e^{-i\vec{p}\cdot\vec{y}} + e^{i\vec{p}\cdot\vec{y}}\right) \frac{\bar{k}^2}{|\vec k|} J^a(\vec{k})  \frac{\bar{p}^2}{|\vec p|} J^b(\vec{p}) \qquad \\
\propto & \frac{1}{\mu^2} f^{bec}f^{ead}\int_{\slashed{k},\slashed{p},\slashed{r}} e^{-\frac{\vec{k}^2}{4\mu^2}}  \frac{\bar{k}}{\bar{p}} J^d(r)  \frac{\bar{k}^2}{|\vec k|} J^a(\vec{k})  \frac{\bar{p}^2}{|\vec p|} J^b(\vec{p})  \nn\\
&  \Big(\frac{p \bar{k}}{\mu^2} ( J^c(\vec p-\vec r+\vec k) + J^c(\vec p-\vec r-\vec k) + J^c(-\vec p-\vec r+\vec k) + J^c(-\vec p-\vec r-\vec k)) \nn\\
& -2 ( J^c(\vec p-\vec r+\vec k) - J^c(\vec p-\vec r-\vec k) + J^c(-\vec p-\vec r+\vec k) - J^c(-\vec p-\vec r-\vec k)) \Big)  
\eE 
This also vanishes for $\mu\to\infty$.

\subsection{Order $e^2J^2$ corrections to the gauge invariant Hamiltonian - 2nd term}
\label{subsec:append3}
We look at the different parts of $\tilde{\Omega}^{(2)}_{ab}(\vec x,\vec v,\vec y)\frac{\delta^2 F^{(0)}}{\delta J^a(\vec v) \delta J^b(\vec y)}$ separately:
\subsubsection{The $\p_y \bar{G}(y-x)$ term}
 \bE{rl}
& \int_{x,v,y}\delta_\mu(\vec x, \vec v) \left(\p_y \bar{G}(y-v)\right) (x-v)^2 (J(\vec v)J(\vec v))_{ab} \frac{\delta^2 F_{GI}^{(0)}}{\delta J^a(\vec y) \delta J^b(\vec x)}  \nn\\
\propto &  f^{cae} f^{deb} \int_{x,v,y} \left(\p_y \bar{G}(y-v)\right) (x-v)^2 J^c(\vec v)J^d(\vec v) \frac{\delta^2 F_{GI}^{(0)}}{\delta J^a(\vec y) \delta J^b(\vec x)}  \\
\propto& f^{cbe} f^{deb} \int_{x,z,v}\delta_\mu(\vec x, \vec v) (x-v)^2 \int_{\slashed{p},\slashed{k}_1,\slashed{k}_2,\slashed{q}} \frac{p}{\bar{p}} e^{i\vec{p}\cdot(\vec{z}-\vec{v})}   J^c(\vec k_1)J^d(\vec k_2) e^{i(\vec{k}_1+\vec{k}_2)\cdot\vec{v}}  \frac{\bar{q}^2}{2|\vec{q}|} \left[e^{-i\vec{q}\cdot\vec{z}} e^{i\vec{q}\cdot\vec{x}} + e^{i\vec{q}\cdot\vec{z}} e^{-i\vec{q}\cdot\vec{x}} \right] \quad\qquad \\
\propto& {1\over\mu^4} \int_{\slashed{k},\slashed{q}} J^a(\vec k)J^a(-\vec k)  \frac{q\bar{q}^3}{|\vec{q}|} e^{-\frac{\vec{q}^2}{4\mu^2}} 
\eE
This vanishes under integration of the angular component of $\vec{q}$.

\subsubsection{The $\bar{G}(y-x)$ term}
\bE{rl}
&\int_{x,v,y}\delta_\mu(\vec x, \vec v) \bar{G}(y-v) f^{ebf} J^e(\vec y) (x-v) f^{cfa}J^c(\vec v) \frac{\delta^2 F_{GI}^{(0)}}{\delta J^a(\vec y) \delta J^b(\vec x)}   \nn\\
\propto &  \int_{\slashed{k},\slashed{q}} J^a(\vec k)J^a(-\vec k)   e^{-\frac{\vec{q}^2}{4\mu^2}} \frac{2\bar{k}\bar{q}^3}{\mu^2|\vec{q}|(\bar{q}^2-\bar{k}^2)}   \\
\propto &  {1\over\mu} \int_{\slashed{k}} \bar{k}^2 J^a(\vec k)J^a(-\vec k)  
\eE
This vanishes for $\mu\to\infty$.

\end{document}